\documentclass[twocolumn,10pt,prb,aps,showpacs,superscriptaddress,preprintnumbers,amsmath,amssymb]{revtex4-1}

\usepackage{graphicx} 
\usepackage{dcolumn} 
\usepackage{bm} 
\usepackage{physics}
\usepackage{amsmath}
\usepackage{ragged2e}

\usepackage{ulem}
\usepackage[]{color}

\usepackage{tabularx}
\newcolumntype{L}[1]{>{\raggedright\arraybackslash}p{#1}} 
\newcolumntype{C}[1]{>{\centering\arraybackslash}p{#1}} 
\newcolumntype{R}[1]{>{\raggedleft\arraybackslash}p{#1}} 

\begin{document}



\title{Determining the local low-energy excitations in the Kondo semimetal CeRu$_4$Sn$_6$ using resonant inelastic x-ray scattering}



\author{Andrea~Amorese}
  \affiliation{Institute of Physics II, University of Cologne, Z{\"u}lpicher Stra{\ss}e 77, 50937 Cologne, Germany}
  \affiliation{Max Planck Institute for Chemical Physics of Solids, N{\"o}thnitzer Stra{\ss}e 40, 01187 Dresden, Germany}
\author{Kurt~Kummer}
  \affiliation{European Synchrotron Radiation Facility, 71 Avenue des Martyrs, CS40220, F-38043 Grenoble Cedex 9, France}
\author{Nicholas~B.~Brookes}
  \affiliation{European Synchrotron Radiation Facility, 71 Avenue des Martyrs, CS40220, F-38043 Grenoble Cedex 9, France}
\author{Oliver~Stockert}
  \affiliation{Max Planck Institute for Chemical Physics of Solids, N{\"o}thnitzer Stra{\ss}e 40, 01187 Dresden, Germany}
\author{Devashibhai~T.~Adroja}
  \affiliation{ISIS Facility, Rutherford Appleton Laboratory, Chilton, Didcot Oxon OX11 0QX, United Kingdom}
\author{Andr\'{e}~M.~Strydom}
  \affiliation{Highly Correlated Matter Research Group, Physics Department, University of Johannesburg, P.O. Box 524, Auckland Park 2006, South Africa}
\author{Andrey~Sidorenko}
  \affiliation{Institute of Solid State Physics, Vienna University of Technology, Wiedner Hauptstra{\ss}e 8-10, 1040 Vienna, Austria}
\author{Hannes Winkler}
  \affiliation{Institute of Solid State Physics, Vienna University of Technology, Wiedner Hauptstra{\ss}e 8-10, 1040 Vienna, Austria}
\author{Diego A. Zocco}
  \affiliation{Institute of Solid State Physics, Vienna University of Technology, Wiedner Hauptstra{\ss}e 8-10, 1040 Vienna, Austria}
\author{Andrey~Prokofiev}
  \affiliation{Institute of Solid State Physics, Vienna University of Technology, Wiedner Hauptstra{\ss}e 8-10, 1040 Vienna, Austria}
\author{Silke~Paschen}  
  \affiliation{Institute of Solid State Physics, Vienna University of Technology, Wiedner Hauptstra{\ss}e 8-10, 1040 Vienna, Austria}
\author{Maurits~W.~Haverkort} 
  \affiliation{Institute for Theoretical Physics, Heidelberg University, Philosophenweg 19, 69120 Heidelberg, Germany}
\author{Liu~Hao~Tjeng} 
   \affiliation{Max Planck Institute for Chemical Physics of Solids, N{\"o}thnitzer Stra{\ss}e 40, 01187 Dresden, Germany}
\author{Andrea~Severing}
   \affiliation{Institute of Physics II, University of Cologne, Z{\"u}lpicher Stra{\ss}e 77, 50937 Cologne, Germany}
   \affiliation{Max Planck Institute for Chemical Physics of Solids, N{\"o}thnitzer Stra{\ss}e 40, 01187 Dresden, Germany}


\date{\today}

\begin{abstract}
We have investigated the local low-energy excitations in CeRu$_4$Sn$_6$, a material discussed recently in the framework of strongly correlated Weyl semimetals, by means of Ce $M_5$ resonant inelastic x-ray scattering (RIXS). The availability of both $^2$F$_\frac{5}{2}$ and $^2$F$_\frac{7}{2}$ excitations of the Ce $4f^1$ configuration in the spectra allows for the determination of the crystal-electric field parameters that explain quantitatively the temperature dependence and anisotropy of the magnetic susceptibility. The absence of an azimuthal dependence in the spectra indicates that all crystal-electric field states are close to being rotational symmetric. We show further that the non-negligible impact of the $\check A_6^0$ parameter on the ground state of CeRu$_4$Sn$_6$ leads to a reduction of the magnetic moment due to multiplet intermixing. The RIXS results are consistent with inelastic neutron scattering (INS) data and are compared to the predictions from \textsl{ab-initio} based electronic structure calculations.
\end{abstract}


\maketitle


In several Ce compounds the localized 4$f$ electrons hybridize with the conduction electrons ($cf$-hybridization) so that hybridization gaps can form and give rise to Kondo insulating, semiconducting or semimetallic ground states.\cite{Riseborough2000} These materials are presently the focus of interest due to the proposal that the combination of strong spin-orbit coupling, bands of opposite parity (4$f$ and 5$d$), plus the hybridization 
induced gap should give rise to strongly correlated non-trivial topological phases.\cite{Dzero2010,Takimoto2011,Dzero2016,Dzsaber2017,Lai93} CeRu$_4$Sn$_6$ is a tetragonal, non-centrosymmetric ($I\bar{4}2m$)\,\cite{Pottgen_1997} compound. Its electrical resistivity increases as temperature decreases which has been attributed to the opening of a hybridization gap of the order of 100\,K.\cite{Das_1992,Strydom_2005,Paschen_2010,Winkler_2012,Guritanu_2013} The absence of magnetic order down to 50\,mK\,\cite{Strydom_2007} and the non-integer valence of 3.08\,\cite{Sundermann2015,Sundermann2017} confirm the importance of strong $cf$-hybridization. Recently, band structure calculations in the LDA+Gutzwiller scheme have suggested that CeRu$_4$Sn$_6$ is a correlated Weyl semimetal,\cite{Xu2017} a conjecture that remains to be tested experimentally, especially since the non-centrosymmetric crystal structure complicates the prediction for gap openings after a band inversion.

To understand the properties of CeRu$_4$Sn$_6$ and to assess the reliability of the theoretical predictions we need to know not only the ground state but also the low-energy excitations of this system. The linear dichroism (LD) in soft x-ray absorption (XAS) and the direction dependence in non-resonant inelastic scattering (NIXS) have shown that the  crystal-electric field (CEF) ground state symmetry must be the $\Gamma_6$\,\cite{Sundermann2015} in agreement with magnetization measurements.\cite{Paschen_2010} However, there is so far no information about the CEF level scheme, i.e. about the energy splittings $\Delta$E$_1$ and $\Delta$E$_2$ and the mixing factor $\alpha$ of the excited CEF  states. The present resonant inelastic x-ray scattering (RIXS) study aims at giving a full description of the CEF level scheme of CeRu$_4$Sn$_6$.

In an ionic model the trivalent (4$f^1$) configuration of Ce is split by the effect of spin-orbit interaction ($\approx$\,280\,meV) in two multiplets, $^2$F$_\frac{5}{2}$  and $^2$F$_\frac{7}{2}$, with 6-fold ($J_z$\,=\,$\left\{-\frac{5}{2};....;+\frac{5}{2}\right\}$) and 8-fold degeneracy ($J_z$\,=\,$\left\{-\frac{7}{2};....;+\frac{7}{2}\right\}$). This degeneracy is further reduced by the interaction with the surrounding ions in the crystal and can be modeled with an effective CEF potential, written as a sum of (renormalized) spherical harmonics $C_k^m=\sqrt{\frac{4\pi}{2k+1}}Y_k^m$:  
\begin{align*}
V_{CEF}(r,\theta,\Phi)=\sum_{k=0}^\infty\sum_{m=-k}^kA_k^mr^kC_k^m(\theta,\Phi)\ .
\end{align*}

The expectation values $\langle r^k \rangle$ cannot be calculated \textsl{ab-initio} and are usually included in the phenomenological CEF parameters $\check A_k^m=A_k^m\langle r^k\rangle$ that must be determined experimentally. Five independent parameters $\check A_2^0$, $\check A_4^0$, $\check A_4^{\pm 4}$, $\check A_6^0$ and $\check A_6^{\pm 4}$ fully describe the CEF problem for a Ce$^{3+}$ ion with tetragonal point symmetry as in  CeRu$_4$Sn$_6$.  Non-zero $\check A_4^{\pm 4}$ and $\check A_6^{\pm 4}$ mix the $J_z$ states  according to $\Delta J_z$\,=\,4, i.e. $J_z$\,=\,$\pm\frac{3}{2}$ and $\mp\frac{5}{2}$, and $J_z$\,=\,$\pm\frac{1}{2}$ and $\mp\frac{7}{2}$, respectively. The intermixing of the two $J$ mulitplets $^2$F$_\frac{5}{2}$ and $^2$F$_\frac{7}{2}$ is usually negligible and the impact of the higher order parameters $\check A_6^0$ and $\check A_6^{\pm 4}$ is small, even on the excited multiplet $^2$F$_\frac{7}{2}$, so that as first approximation the three $A_k^m$ parameters with $k=2$ and $4$ describe the CEF problem. The three Kramers doublets of $^2$F$_\frac{5}{2}$ can then be written in the well known $\ket{J,\pm J_z}$ form as 
$\Gamma_7^1$\,=\,$\alpha \ket{\frac{5}{2};\pm \frac{5}{2}}\,+\,\lvert \sqrt{1\,-\,\alpha^2} \rvert \ket{\frac{5}{2};\mp \frac{3}{2}}$, 
$\Gamma_7^2$\,=\,$\lvert \sqrt{1\,-\,\alpha^2} \rvert \ket{\frac{5}{2};\pm \frac{5}{2}}\,-\,\alpha \ket{\frac{5}{2};\mp \frac{3}{2}}$, 
and $\Gamma_6$\,=\,$\ket{\frac{5}{2};\pm \frac{1}{2}}$. 

\begin{figure}
 \includegraphics[width=0.98\columnwidth]{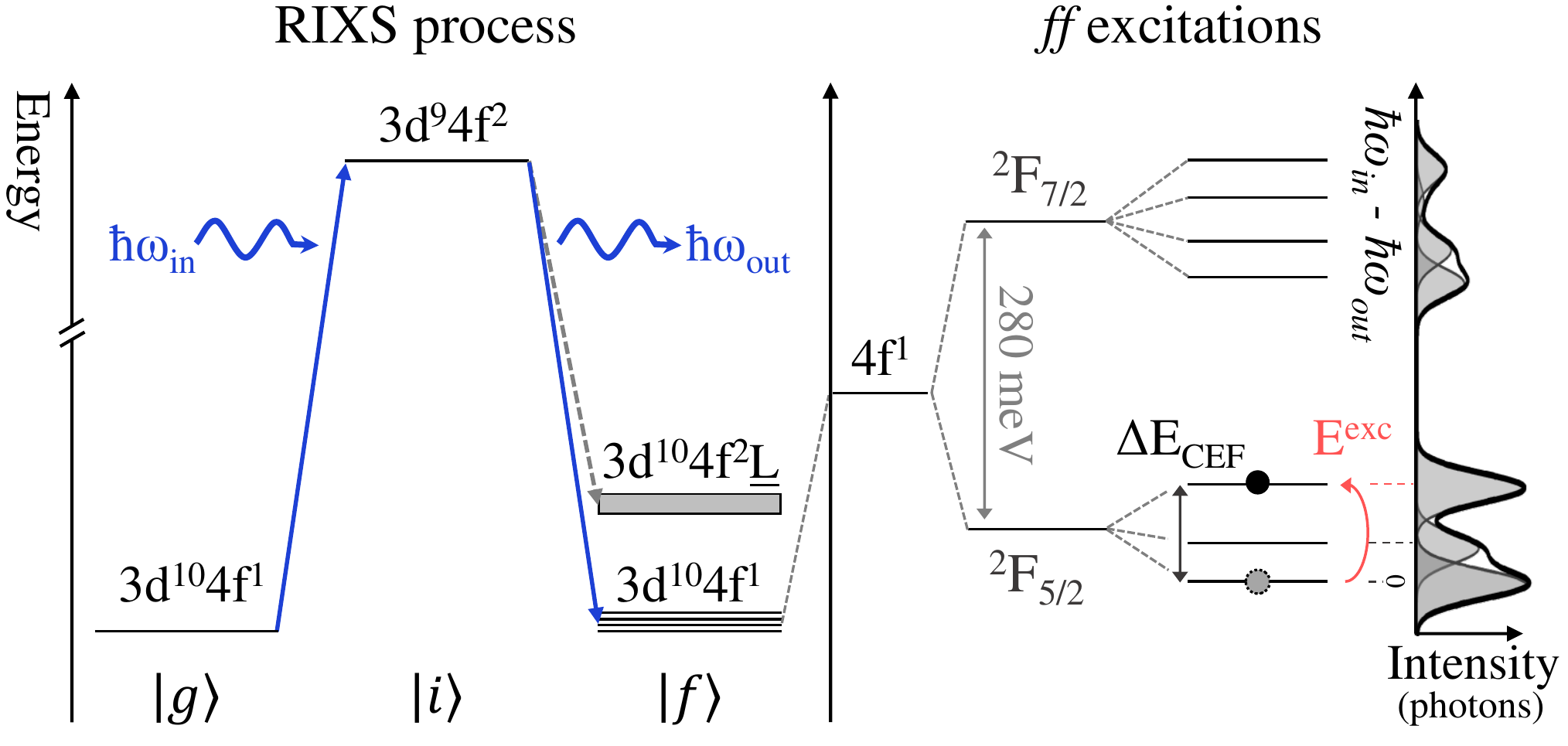}
 \caption{Cerium $M_{4,5}$-edge RIXS process and $ff$ excitations, see text.}  
\label{Fig:RIXS&ff}
 \end{figure}

We apply high resolution soft x-ray RIXS, an innovative spectroscopic technique, for determining the CEF level scheme of CeRu$_4$Sn$_6$. First feasibility experiments have proven its sensitivity to $ff$ excitations.\cite{Amorese_2016,Amorese_2018} Following a second order perturbation treatment, RIXS can be interpreted as the absorption of a photon resonant at a core edge of an ion in the system, followed by a re-emission. When the system is left in an excited state, excitation energies are detected as energy losses of the scattered photons. This is depicted in in Fig.\,\ref{Fig:RIXS&ff} for the RIXS process at the Ce $M_{4,5}$ edge (3$d$\,$\rightarrow$\,4$f$). From the ground state $\ket{g}$ a 3$d$ electron is excited into the 4$f$ shell (intermediate state $\ket{i}$) and then decays into the final state $\ket{f}$ that can be the ground state (elastic scattering) or an excited state of the same configuration (magnons, phonons, $ff$ excitations), or a different configuration via charge transfer.\cite{Nakazawa_1996, Butorin_1996, Dallera_2001} Charge transfer excitations are usually broad and featureless compared to the (nearly) resolution limited $ff$ excitations and only contribute with a widely spread background, usually at higher energy losses. We neither expect collective magnon nor phonon excitations in CeRu$_4$Sn$_6$ due to the absence of magnetic order and the likely low electron-phonon coupling of the $4f$ subshell. This is an effect of the resonant process, which imposes, in addition to element and valence selectivity, that all observable excitations must be coupled with the electronic levels involved in the RIXS process.\cite{Ament_2011b} The strong resonance process ensures that the electronic excitations overwhelm dramatically (if not completely) the complex phonon background that usually is more visible in inelastic neutron scattering (INS) experiments. In addition, the very favorable signal-to-noise ratio in comparison to INS and the ability to focus x-rays allows measuring very small single crystals (surface $\ll 1$\,mm$^2$).

Specific selection rules for polarization and scattering geometries provide further information about the magnetic versus charge origin of excitations, symmetry of the ground and excited states, and their orientation in the unit cell even in presence of higher than twofold rotational symmetry.\cite{Braicovich_2010b, Moretti_2011, Minola_2015, Braicovich_2014,Amorese_2018} The latter is due the fact that the selection rules in RIXS are $\Delta J_z$\,=\,0,\,$\pm1$,\,$\pm2$, i.e. RIXS is not dipole limited as INS with $\Delta J_z$\,=\,0,\,$\pm1$.  The RIXS spectrum should therefore provide a background free mapping of the Ce$^{3+}$ $4f$ energy levels, providing a direct measure the CEF splitting of both $^2$F$_\frac{5}{2}$ and $^2$F$_\frac{7}{2}$, as depicted in Fig.\,\ref{Fig:RIXS&ff}.

\begin{figure*}
\includegraphics[width=0.98\columnwidth]{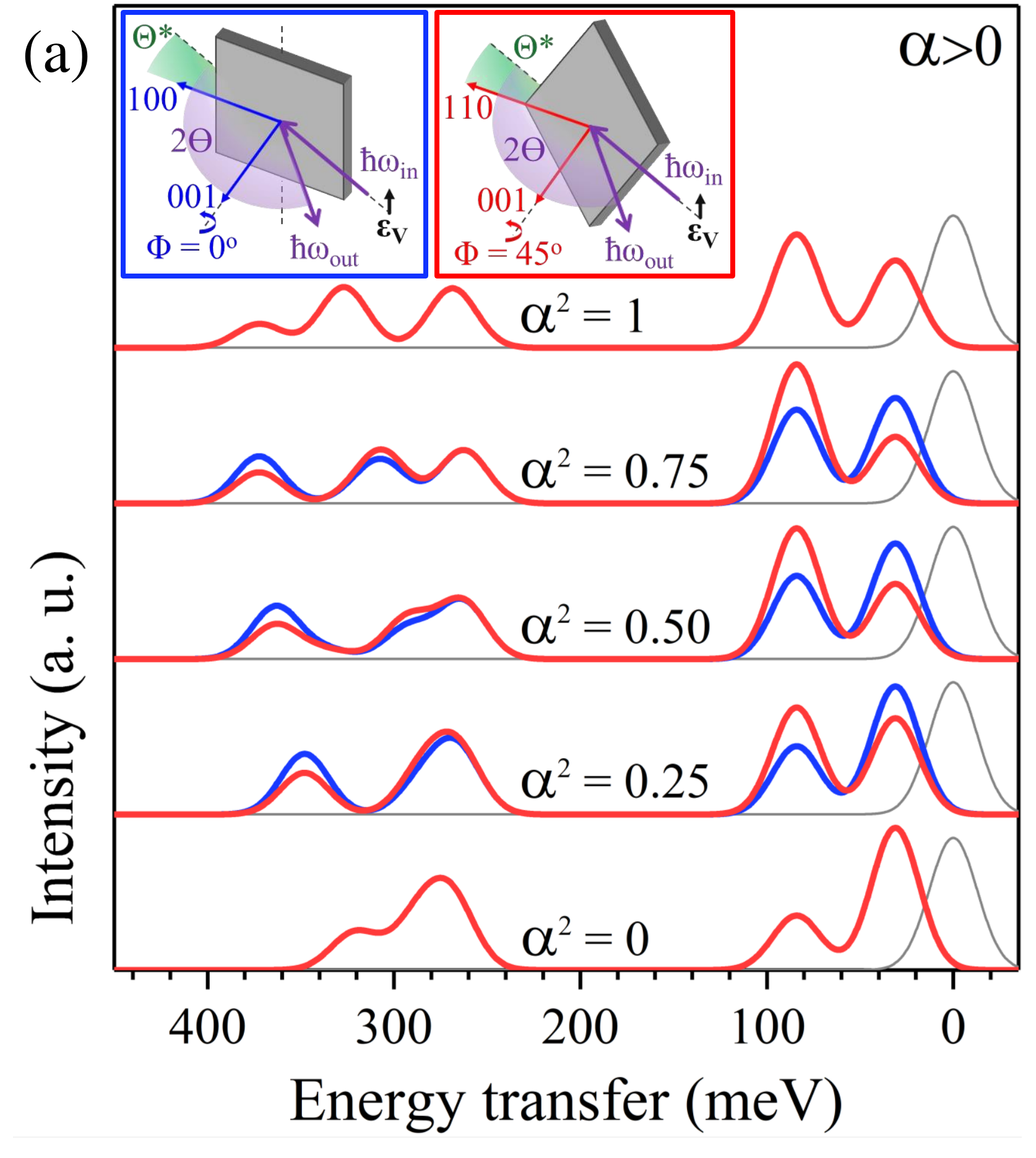}
\includegraphics[width=0.98\columnwidth]{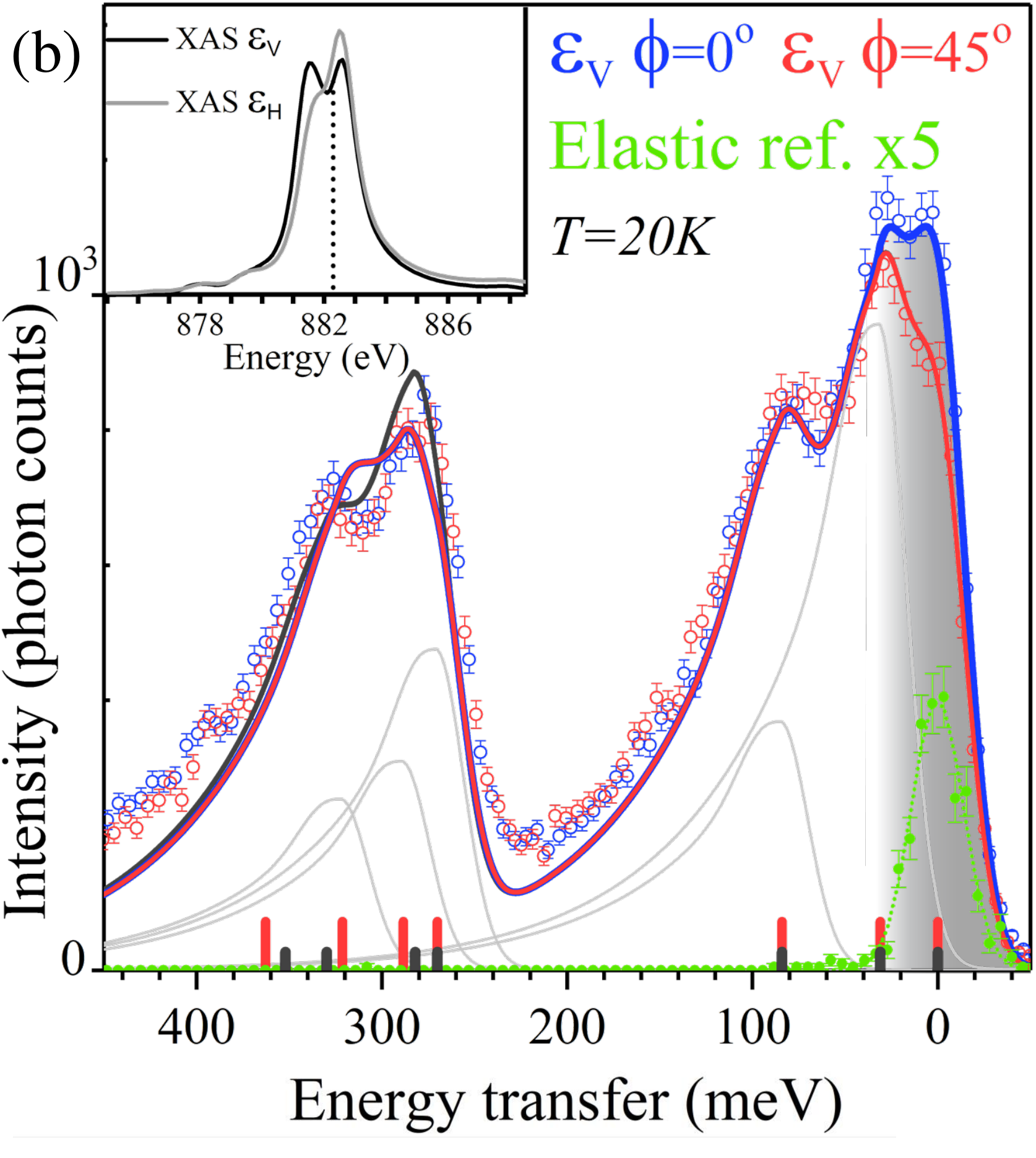}
\caption{(a) Calculated RIXS spectra at $T$\,=\,20\,K as a function of the mixing parameter $\alpha^2$ for $\alpha > 0$, with vertical polarization, for the two geometries $\phi=0^\circ$ and $\phi=45^\circ$ (see inset). (b) Experimental RIXS data (circles) with statistical error bars with the same scattering geometry as in the calculated RIXS spectra on the right. The red and blue lines (black line) show the full multiplet simulation with three (four) crystal-field parameters (the fourth being$\check A_6^0$\,=\,200\,meV) using an asymmetric lineshape for the CEF excitations, see red (black) ticks and gray lines. The gray shading shading indicates the elastic region, see text. The elastic reference (green) shows the Gaussian response function of the beamline. XAS M$_5$ edge and incident energy (dotted line) shown in inset.}  
\label{Fig:Exp}
\end{figure*}

$ff$ excitations are intra-atomic and well localized so that they can be simulated with a single ion full-multiplet calculation. Figure\,\ref{Fig:Exp}(a) shows the simulations of RIXS photon energy loss spectra at the Ce $M_{5}$ edge for a single crystal, performed with the full-multiplet code Quanty.\cite{Haverkort2016,Haverkort_2012} The atomic 4$f$-4$f$ and 4$d$-4$f$ Coulomb interactions were calculated with Cowan's atomic structure code.\cite{Cowan} Typical reductions of 20\,\% and 30\,\%\,\cite{Hansmann2008,Sundermann2015} have been applied, respectively, to account for configuration interaction processes not included in the Hartree-Fock scheme.\cite{Tanaka1994} The spin-orbit interaction in the $4f$ shell has been reduced by $10$\,\%.\cite{Amorese_2016, Amorese_2018} The calculations were set up for the back-scattering geometry ($2\theta=150^\circ$), grazing incidence ($\Theta^*=20^\circ$), and with the tetragonal $c$-axis (normal to the sample surface) in the scattering plane (see insets in Fig.\,\ref{Fig:Exp}(a)). The calculations were carried out for the vertical polarization ($\epsilon_v$) and two different sample orientations; blue lines for the [100] ($\phi$\,=\,$0^\circ$) and red lines for the [110] ($\phi$\,=\,$45^\circ$) in the scattering plane. The calculated intensities are convoluted with a 30\,meV FWHM Gaussian function representing the resolution achievable in experiments.  The elastic intensity cannot be reliably calculated since surface roughness as well as low energy collective excitations contribute to the (quasi)elastic signal. Instead we show a resolution function of arbitrary height (gray lines centered at 0\,meV). 

The calculations in Fig.\,\ref{Fig:Exp}(a) are for a ground state with $\Gamma_6$ symmetry\,\cite{Sundermann2015} and for several mixing factors $\alpha$ of the excited states at fixed energy splittings $\Delta$E$_1$ and $\Delta$E$_2$. $\alpha$ is defined such that the bottom curve ($\alpha$\,=\,0) refers to the sequence $\ket{\frac{5}{2};\pm \frac{1}{2}}$, $\ket{\frac{5}{2};\pm \frac{3}{2}}$ and $\ket{\frac{5}{2};\pm \frac{5}{2}}$ for the ground, first and second excited state. Accordingly, $\alpha$\,=\,1 resembles the sequence $\ket{\frac{5}{2};\pm \frac{1}{2}}$,$\ket{\frac{5}{2};\pm \frac{5}{2}}$ and $\ket{\frac{5}{2};\pm \frac{3}{2}}$. Figure\,S1 in the \textsl{Supplemental Material} shows calculations for different ground states and excited state sequences.\cite{Supp}  The calculations are based on only three parameters, $\check A_2^0$, $\check A_4^0$, and $\check A_4^{\pm 4}$. Note that RIXS is sensitive to the orientation of the orbitals in the unit cell: changing the sign of $\check A_4^{\pm 4}$, which is equivalent to changing the sign of $\alpha$ in the calculation (not shown), causes a $45^\circ$ rotation of the orbitals in the $ab$ plane,\cite{WillersPRL109} and correspondingly inverts the $\Phi$ dependence of the spectra (exchanges the blue and red lines).

The RIXS experiment on single crystalline CeRu$_4$Sn$_6$\,\cite{Paschen_2010,Winkler_2012,Prokofiev_2012} was preformed at the ERIXS spectrometer of the ID32 beamline at ESRF, Grenoble, France with a resolution of 30\,meV at the Ce $M_5$-edge ($\approx 880$\,eV).\cite{Amorese_2018} The inset of Fig.\,\ref{Fig:Exp}(b) shows the $M$-edge XAS spectra measured in the same geometry pointing out the incident photon energy (882.2\,eV) used for the RIXS spectroscopy. Figure\,\ref{Fig:Exp}(b)) shows data for the two sample orientations $\phi$ as in the calculation, i.e. $\phi$\,=\,$0^\circ$ (blue dots) and $\phi$\,=\,$45^\circ$ (red dots). Both spectra have been acquired with incident vertical polarization. Other spectra acquired with different experimental settings can be found inRef\,\onlinecite{Supp}. The green dots are the measurements of carbon tape that serves as an elastic reference. More details of the beamline and set-up can also be found in Ref\,\onlinecite{Supp}.

The experimental RIXS spectra show two groups of peaks, the first one at 0\,-\,100\,meV and the second at 250\,-\,400\,meV corresponding to the $^2$F$_\frac{5}{2}$ multiplet and $^2$F$_\frac{7}{2}$ multiplet, respectively. The spectra of both sample orientations show the expected three peaks in the $^2$F$_\frac{5}{2}$ energy range: the elastic peak ($E_0=$0\,meV) plus two inelastic peaks at about $\Delta$E$_1$\,$\approx$\,30\,meV and $\Delta$E$_2$\,$\approx$\,85\,meV. In the $^2$F$_\frac{7}{2}$ energy range we would expect to see four excitations due to the splitting into four Kramers doublets, however, while they are intense, they also seem to be too close in energy to be resolved.

The inelastic signals of the two experimental spectra are superimposed, reflecting no dependence on the rotation $\phi$ around the $c$ axis, thus suggesting the orbitals must have, or are close to, rotational symmetry which occurs in the presence of pure $J_z$ states (see e.g. Ref.\,\onlinecite{Hansmann2008}). Thus, the intermixing of $J_z$\,=\,$\pm$3/2\,and\,$\mp$5/2 and of $J_z$\,=\,$\pm$1/2\,and\,$\mp$7/2 of the excited states is next to zero, i.e. $\check A_4^{\pm 4}$ and also $\check A_6^{\pm 4}$ must be negligibly small.

\begin{figure}
 \includegraphics[width=0.98\columnwidth]{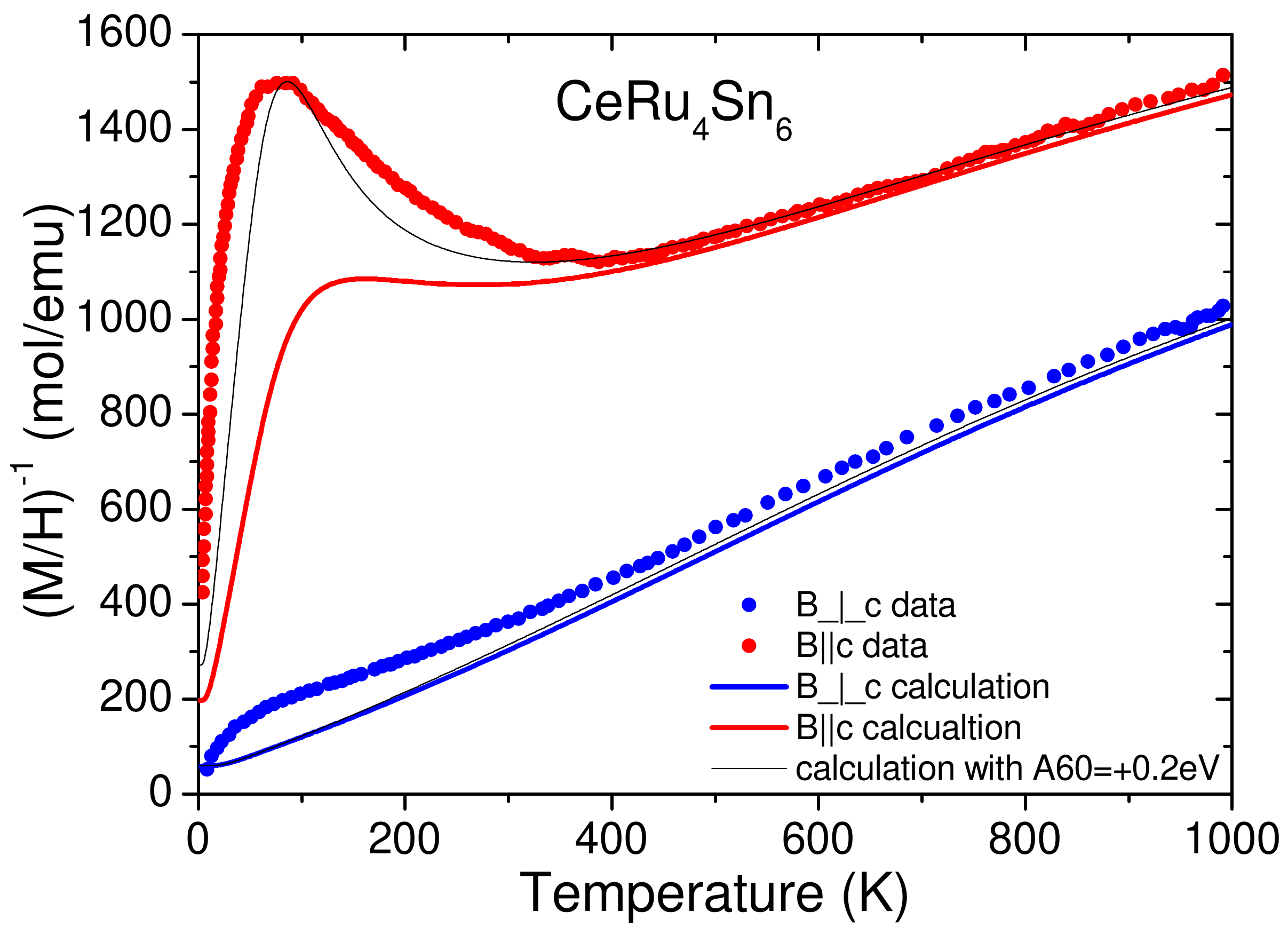} 
 \caption{$(M/H)^{-1}$ curves of CeRu$_4$Sn$_6$ measured with a 5\,T and 6\,T magnetic field parallel (red dots) 
and perpendicular (blue dots) to the tetragonal $c$ axis for $T$\,$\leq$300\,K and $T$\,$\geq$300\,K, respectively. 
The blue and red lines are the CEF-only calculation with $\check A_6^0$\,=\,0 and the black lines with 
$\check A_6^0$\,=\,200\,meV.}
 \label{chi}
 \end{figure}

\begin{figure}
 \includegraphics[width=0.98\columnwidth]{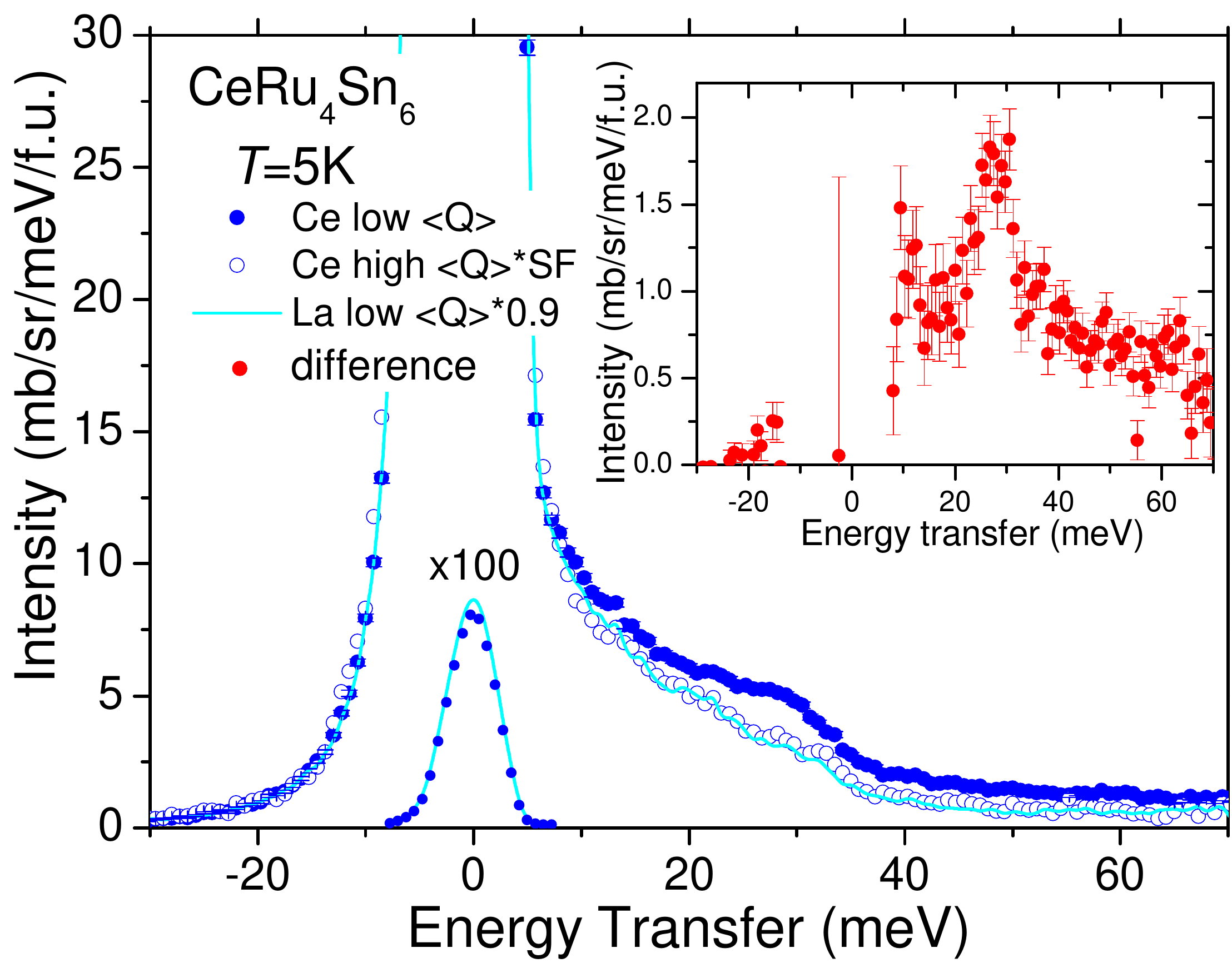}
 \caption{Inelastic neutron scattering (INS) data at 5\,K of CeRu$_4$Sn$_6$ (blue dots) at low momentum transfers 
($<$Q$>$\,=\,2.2\,\AA$^{-1}$) compared to the nuclear scattering of high to low $<$Q$>$ scaled CeRu$_4$Sn$_6$ 
(open dots) and non-magnetic scattering-length corrected low $<$Q$>$ LaRu$_4$Sn$_6$ (cyan line) 
(see Supp. Mat.\,\onlinecite{Supp}). The strong elastic scattering has been divided by a factor of 100. Inset: magnetic 
scattering as determined form the difference of low $<$Q$>$ CeRu$_4$Sn$_6$ and high to low $<$Q$>$ scaled 
CeRu$_4$Sn$_6$ nuclear scattering. All data are normalized to absolute intensities.}
 \label{INS}
\end{figure}

In the following we will compare the experimental data with calculations. The absence of a pronounced $\phi$ dependence  in the experiment, see Fig.\,\ref{Fig:Exp}(b), indicates that $\alpha$ is very close to 0 or 1, see  Fig.\,\ref{Fig:Exp}(a). Taking further into account the intensity ratios of  the main peaks in either multiplet, we observe that the $\alpha$\,$\approx$\,0 calculation shows better resemblance of the  experimental spectra. The only pronounced deviation in the experimental spectra between the two sample rotations is in the  elastic region (see gray shaded area). This could have either the trivial reason that the quality of the sample surface is  different in the two positions, or it shows that $\alpha$ is positive although close to zero (see Fig.\,\ref{Fig:Exp}(a)). We can conclude that  the excited CEF states of the lower multiplet are almost pure $\ket{J_z}$ states with $\ket{\frac{5}{2};\pm \frac{1}{2}}$ being the ground state, $\ket{\frac{5}{2};\pm \frac{3}{2}}$ the first excited state  and $\ket{\frac{5}{2};\pm \frac{5}{2}}$ the second excited state. 

Although we find a qualitative agreement between data and calculations, the overall spectral shapes differ.  The calculation in Fig.\,\ref{Fig:Exp}(a) was performed with a Gaussian broadening resembling the resolution function but the experimental spectral shapes are clearly non-Gaussian. Ignoring the non-Gaussian lineshape  could lead to larger energy transfers and/or unaccounted spectral weights.\cite{Amorese_2018} We therefore  used an empirical asymmetric lineshape for the $ff$ excitations (see Ref.\,\onlinecite{Supp}). The red (blue) line in  Figure\,\ref{Fig:Exp}\,(b) is the result of a calculation where each excitation is treated with the same,  asymmetric lineshape and only the elastic peak is still convoluted with the Gaussian resolution function. We find the $\ket{\frac{5}{2};\pm \frac{3}{2}}$ state at $\Delta E_1$\,=31\,meV and the  $\ket{\frac{5}{2};\pm \frac{5}{2}}$ at $\Delta E_2$\,=84\,meV with parameters  $\check A_2^0$\,=\,-163.7\,meV $\check A_4^0$\,=\,-13.6\,meV and  $\check A_4^4$\,=\,0\,meV.  Energies and excited state sequences of the $^2$F$_\frac{7}{2}$ multiplet are given in Ref\,\onlinecite{Supp}.  The red ticks at the bottom scale denote the positions of the CEF excitations and the gray lines resemble  the actual intensities. Note that in the calculation, the highest transition has zero intensity because it corresponds  to a $\Delta J_z$\,=\,3 ($\ket{J_z=\pm \frac{1}{2}}\rightarrow \ket{J_z=\pm\frac{7}{2}}$) transition which  is not accessible due to selection rules. Also, spectra acquired with different experimental settings confirm  the aforementioned results (see Ref.\,\onlinecite{Supp}).

Having determined the CEF parameters from the RIXS data, we now investigate to what extend these findings can explain the magnetic properties of the material. Figure\,\ref{chi} shows the experimental inverse magnetic susceptibility $(M/H)^{-1}$ as a function of temperature for the magnetic field parallel (red dots) and perpendicular (blue dots) to the tetragonal $c$-axis. We have calculated $(M/H)^{-1}$ using the CEF parameters extracted from the RIXS experiment and plotted the results also in Fig.\,\ref{chi}, for magnetic fields parallel (red lines) and perpendicular (blue lines) to the $c$-axis. This three-parameter CEF model reproduces $(M/H)^{-1}$ very well at temperatures above 400\,K.

We now consider the higher order parameters $\check A_6^0$ and $\check A_6^4$. As stated above, $\check A_6^4$ must be, like $\check A_4^4$, close to zero. Introducing a positive non-zero $\check A_6^0$ leads to an increase in the peak intensity ratio in the energy range of the $^2$F$_\frac{7}{2}$ multiplet. The black line and black ticks in Fig.\,\ref{Fig:Exp}(b) are the result of a calculation with $\check A_6^0$\,=\,200\,meV. $\check A_6^0$ does not affect the high temperature anisotropy of $(M/H)^{-1}$ but improves the agreement of CEF-only calculation and data for fields parallel $c$ at low temperatures (see black lines in Fig.\,\ref{chi}). The non-zero $\check A_6^0$ increases the multiplet intermixing so that the new ground state wave function $|GS\rangle$ contains now a considerable amount of the $^2$F$_\frac{7}{2}$ multiplet, 
\begin{align*}
|GS\rangle=0.99|J=5/2,J_z=\pm\frac{1}{2}\rangle+0.12|J=7/2,J_z=\pm\frac{1}{2}\rangle.
\end{align*}
This goes a long with a reduction of the magnetic moment $\mu_{\|c}$ from 0.45 to 0.33\,$\mu_B$ as calculated from the CEF model . The further impact of $\check A_6^0$ on the excited states is listed in Table S1 in Ref\,\onlinecite{Supp}. 

The classical tool for determining the CEF scheme in rare earth compounds is INS so that compatibility of RIXS and INS data is essential. Figure\,\ref{INS} shows time-of-flight INS data of polycrystalline CeRu$_4$Sn$_6$ on an absolute intensity scale. Polycrystalline LaRu$_4$Sn$_6$ was measured as non-magnetic reference compound. The low angle data are grouped for low momentum transfers $<$Q$>$\,=\,2.2\,\AA$^{-1}$ where the magnetic form factor is large. The spectra contain incoherent nuclear elastic and inelastic (phonon) scattering as well as, in case of Ce, incoherent magnetic scattering (see full blue dots). To extract the magnetic scattering, the nuclear contribution has been assigned by high to low $<$Q$>$ scaling (open dots) and scaling of the non-magnetic reference data (cyan line). More experimental details as well as explanations of the phonon correction are given in Ref.\,\onlinecite{Supp}. The difference of full and open dots yields the magnetic scattering (see inset).  There is a clear peak at about 30\,meV in agreement with RIXS. Its magnetic origin has been further confirmed by comparing the Ce and La data at large momentum transfers (see Fig.\,S3\,(a)) in Ref.\,\onlinecite{Supp}). The second CEF excitation at 80 to 85\,meV is not only outside the energy window of the present INS experiment, it is also dipole forbidden ($\Delta J_z$\,=\,$\pm$2), i.e. not observable in an INS experiment. The integrated intensity of the excitation at 30\,meV is compatible with the CEF model that describes the RIXS data, at 5\,K as well as at 300\,K (see Ref.\,\onlinecite{Supp}). It is important to mention that high resolution INS data 
(not shown) confirm the absence of any lower lying CEF excitation.

We now compare the RIXS and INS results with electronic structure calculations. Wissgott \textsl{et al.}\,\cite{Wissgott2016} performed density functional theory (DFT) plus dynamical mean field theory (DMFT) calculations of CeRu$_4$Sn$_6$ for treating the strong correlation effects of $f$-electrons of Ce and find a $\ket{\frac{5}{2};\pm \frac{1}{2}}$ ground state with some mixing of the higher multiplet $^2$F$_\frac{7}{2}$ and with some contribution of $\ket{\frac{5}{2};\pm \frac{3}{2}}$ due to $cf$-hybridization.  Xu \textsl{et al.}\,\cite{Xu2017} performed LDA\,+\,Gutzwiller calculations and find the same CEF ground state as Wissgott \textsl{et al.}, but a stronger contribution of $\ket{\frac{5}{2};\pm \frac{5}{2}}$than $\ket{\frac{5}{2};\pm \frac{3}{2}}$ mixed in by hybridization. In RIXS, we find the $\ket{\frac{5}{2};\pm \frac{3}{2}}$ state at about 30\,meV and the $\ket{\frac{5}{2};\pm \frac{5}{2}}$ above 80\,meV. Hence, the Wissgott \textsl{et al.} scenario seems to be closer to the experiment.  

In NIXS and XAS hybridization effects were seen in the reduction of the ground state dichroism.\cite{Sundermann2015} In RIXS, it should show in a modified elastic or quasielastic signal but in the elastic region RIXS is resolution limited and also hampered by surface effects. However, the strongly asymmetric lineshape in the RIXS spectra are a clear remnant of the strong $cf$-hybridization: the localized 4$f$ states states decay into the continuum. Presently, these effects cannot be calculated quantitatively, so that in the analysis the lineshape had to be treated empirically. 

In summary, the soft RIXS study of CeRu$_4$Sn$_6$ yields a CEF potential with a $\Gamma_6$ ground state and $\Gamma_7$ states at about 30 and 85\,meV with a mixing factor $\alpha$\,$\approx$\,0 that reproduces the high temperature anisotropy of $(M/H)^{-1}$; no adjustment of energy transfers of mixing parameters was required to obtain the excellent agreement. The introduction of the higher order CEF parameter $\check A_6^0$ even reproduces the peak in $M/H^{-1}$ at about 60\,K for fields parallel to $c$ by reducing the magnetic moment $\mu_{\|c}$ of the ground state via intermultiplet mixing. It might well be that here the $J$ mixing mimics to some extend the reduction of the ground state moment due to the presence of strong $cf$-hybridization. The latter shows up in the RIXS spectra as a strongly asymmetric lineshape.

\section{Acknowledgment}
A.A. and A.S. benefited from financial support of the Deutsche Forschungsgemeinschaft through grand SE1441/4-1. A.P., H.W., A.S., D.A.Z., and S.P. acknowledge financial support from the Austrian Science Fund (FWF I 2535-N27) and S. Stojanovic for assistance in the sample preparation. A.M.S. thanks the SA NRF (93549) and the URC/FRC of UJ

\section{Appendix}
 
\subsection{RIXS calculations for different order of states}

\begin{figure}[ht]
 \includegraphics[width=0.99\columnwidth]{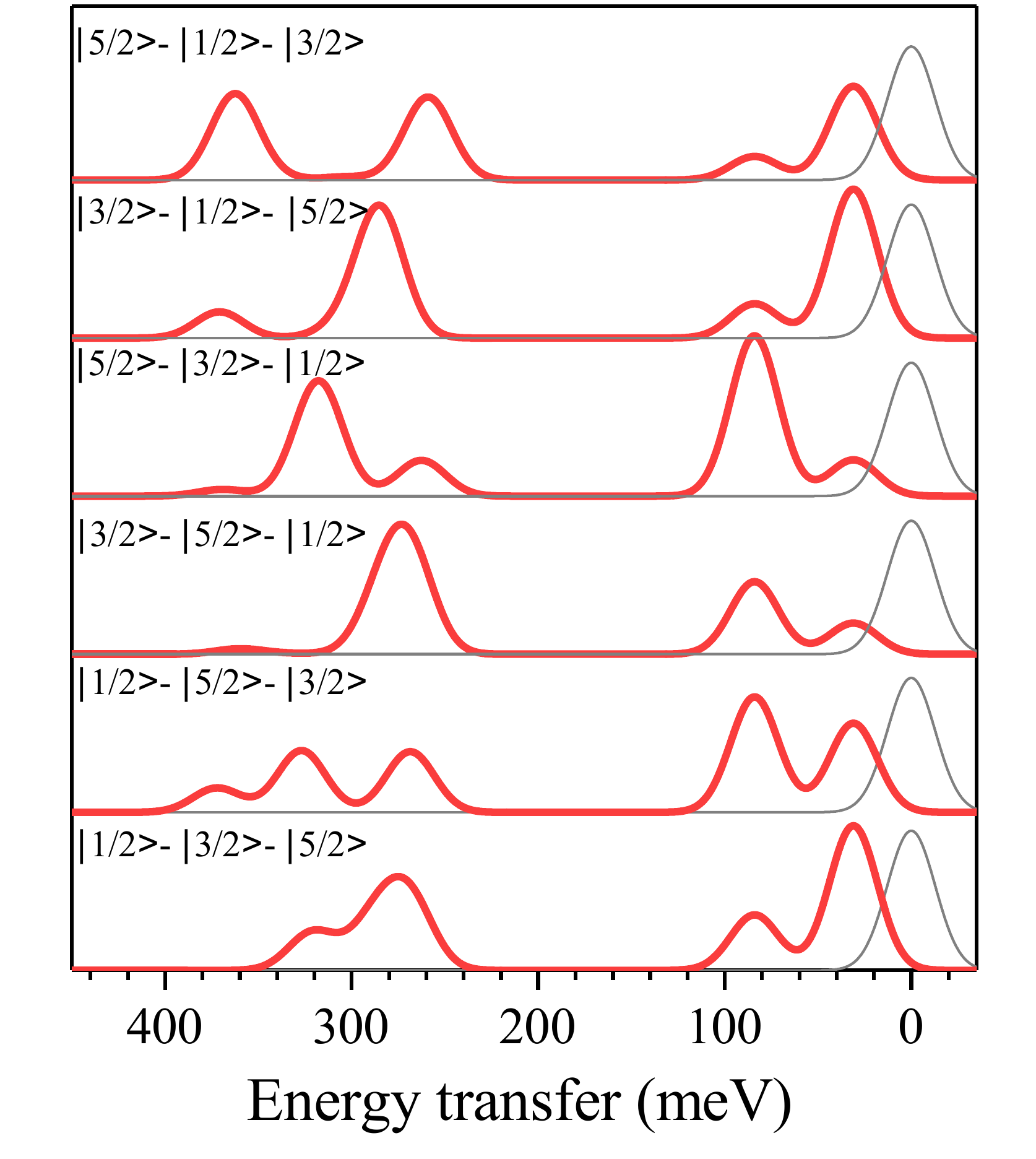}
 \justify
 FIG.\,S1\,\,(color online) RIXS calculations for different pure $J_z$ crystal field ground states and order of states in the ground multiplet.
\label{Fig:/Other_Jz}
 \end{figure} 

Figure\,S1 shows calculated RIXS spectra for energy splitting of 31 and 84\,meV assuming different pure $J_z$ ground states and order of excited states. Only the peak intensity ratio of the the sequence $\ket{\frac{5}{2};\pm \frac{1}{2}}$, $\ket{\frac{5}{2};\pm \frac{3}{2}}$, and $\ket{\frac{5}{2};\pm \frac{5}{2}}$ agrees with the experimental data in both the $^2$F$_{\frac{5}{2}}$ and $^2$F$_{\frac{7}{2}}$ multiplets.  The calculations have been performed with the full multiplet code Quanty\,\cite{Haverkort2016,Haverkort_2012}, using the atomic values from Cowan's code\,\cite{Cowan} and reduction factors as given in the main text.
 
\subsection{RIXS beamline and set-up}
The RIXS experiment was preformed at the ERIXS spectrometer on the ID32 beamline at ESRF, Grenoble France with a resolution of 30\,meV at the Ce $M_5$-edge ($\approx 880$\,eV)\,\cite{Amorese_2018}. The optimal combined energy resolution of the beamline and spectrometer was obtained by using the 1600\,lines/mm grating of the VLS-PGM monochromator and the 2500\,lines/mm grating of the spectrometer. The spectra were acquired with an Andor iKon-L CCD detector using a single photon centroid elaboration method \cite{Amorese_2018a,Kummer2017} in order to overcome the spatial resolution limits of the detector and completely remove the background produced by the darkcurrent and readout noise of the CCD. The acquisition time is slightly more than 4 hours for each spectrum. The instrument 30\,meV-FWHM Gaussian response function was estimated by a 10 minute acquisition of the elastic non-resonant scattering of a carbon tape. 

\subsection{Line shape in RIXS}
The line shape was chosen empirically in order to account for asymmetry of the spectral response of the $ff$ excitations. With the aim of using the lowest number of free parameters for the new lineshape $L$, we modified the resolution limited Gaussian response functions $G$ by using an exponential tail function $T$. 

\begin{equation} \label{eq:Lineshape}
L(E)=G(E)-(1-G(E))\times T(E,P)
\end{equation}
with
\begin{equation} \label{eq:Tail}
T=
 \begin{cases}
      e^{-\frac{E}{P}} &\text{if}\ E>0 \\
      0 &\text{if}\ E\leq 0 
   \end{cases} \text{ ,}
\end{equation}
 $P$ being the only asymmetry parameter, which was set to $70$\,meV. This tail was not applied to the elastic peak, which is mostly produced by the diffuse scattering, sample surface roughness and other phenomena not connected to the electronic interactions at the origin of the additional intensity in the experimental spectra. 

\subsection{RIXS spectra plus simulation for different experimental configurations}
\begin{figure*}[ht]
  \includegraphics[width=1.8\columnwidth]{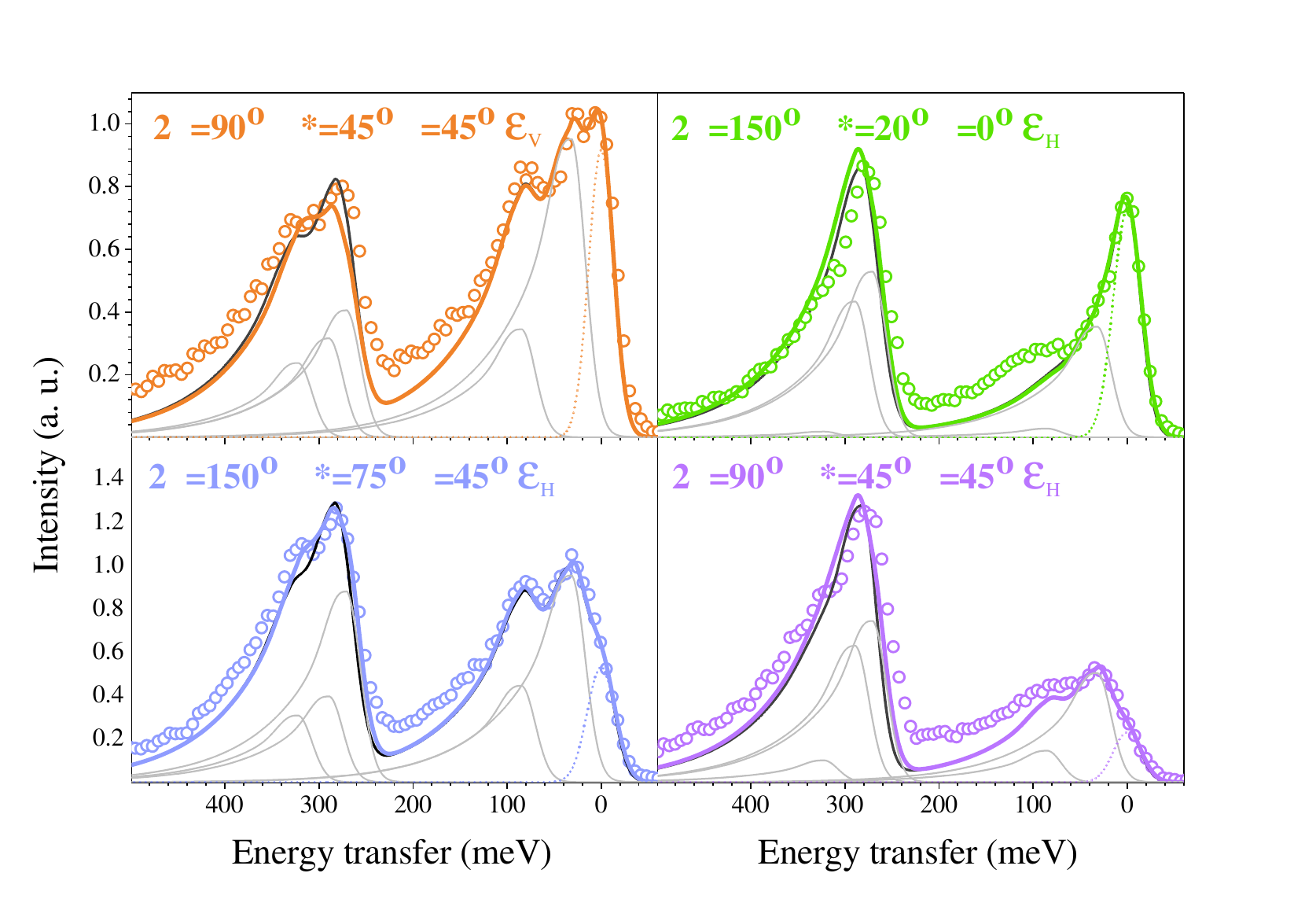}
\justify
 FIG.\,S2\,\,RIXS spectra of CeRu$_4$Sn$_6$ (circles) for different scattering geometries and polarizations as given in figures. Coloured lines resemble the calculated spectra based on a three CEF parameters calculation, black lines considering $\check A_6^0$\,=\,200\,meV.
 \end{figure*}

Figure\,S2 shows experimental RIXS spectra for scattering angles $\theta$, sample rotations $\Theta^*$ and polarizations $\epsilon_V$ (vertical) and $\epsilon_H$ (horizontal). The solid lines are the result of the full multiplet CEF calculations with the asymmetric line shape. The coloured lines are based on the three paramater ($\check A_2^0$, $\check A_4^0$ $\check A_4^4$) calculation, the black lines are the result with non-zero $\check A_6^0$. The dotted colored lines at zero energy transfer are 30\,meV Gaussian resolution functions and the gray lines represent the spectral weights of the CEF excitations. 

\subsection{Inelastic neutron scattering experiment}
For the inelastic neutron scattering experiment polycrystalline\,\cite{Pottgen_1997} CeRu$_4$Sn$_6$ and LaRu$_4$Sn$_6$ samples were used. The experiment was performed at the time-of-flight spectrometer HET at ISIS, Rutherford Laboratory using incident energies of 80\,meV and 11\,meV (not shown) with energy resolutions of 4.2\,meV and 0.6\,meV full (FWHM), respectively, at zero energy transfer. For the data with 80\,meV incident energy the grouping of the low angle banks results in an averaged momentum transfer $<$Q$>$\,=\,2.2\,\AA$^{-1}$ and of the high angle banks to $<$Q$>$\,=\,9.98\,\AA$^{-1}$. All data are normalized to absolute cross-sections, making the comparison with CEF calculations meaningful because this way intensities as well as energy positions have to be reproduced. 

For extracting the magnetic scattering, the nuclear scattering has been assigned by two methods: 1) The high $<$Q$>$ cerium data have been scaled to low $<$Q$>$ (see open dots in Fig.\,4 in the main text) with a scaling factor $SF$ as function of energy that has been determined by dividing the high and low $<$Q$>$ La reference data. 2) The low angle La data have been multiplied with the factor 0.9 in order to account for the difference in the averaged nuclear scattering (cyan line in Fig.,4 in the main text). Both methods yields the same phonon contribution within our accuracy. No statement can be made about the quasielastic magnetic scattering because of the strong nuclear elastic scattering in combination with limited resolution.

Figure\,S3\,(a) shows INS data of CeRu$_4$Sn$_6$ and of non-magnetic LaRu$_4$Sn$_6$ (times 0.9) at $T$\,=\,5\,K and at high angles where the nuclear scattering dominates. These nuclear data do not show an excess of intensity in the cerium data at about 30\,meV, thus confirming the magnetic origin of the 30\,meV peak in the low angle data in Fig.\,4 in the main text.

Figure\,S3\,(b) shows the low angle INS data at room temperature. There is a slight shift of magnetic intensity towards higher energy transfers. This is expected because at 295\,K the first excited state is partially populated so that excitation from the first to the second state are possible; at 5\,K the calculations gives for the excitations at 30\,meV a cross-section of 1.78\,barn, at 295\,K only 1.34\,barn but additional 0.25\,barn for the transition from the first to the second excited state. Given the large error bars of the difference spectra that resemble the magnetic scattering we can state that the total intensity in the accessible energy window of 10 to 70 meV should remain unchanged, with a slight shift of intensity towards larger energy transfers. This agrees with the observation.

 \begin{figure*}
\includegraphics[width=0.9\columnwidth]{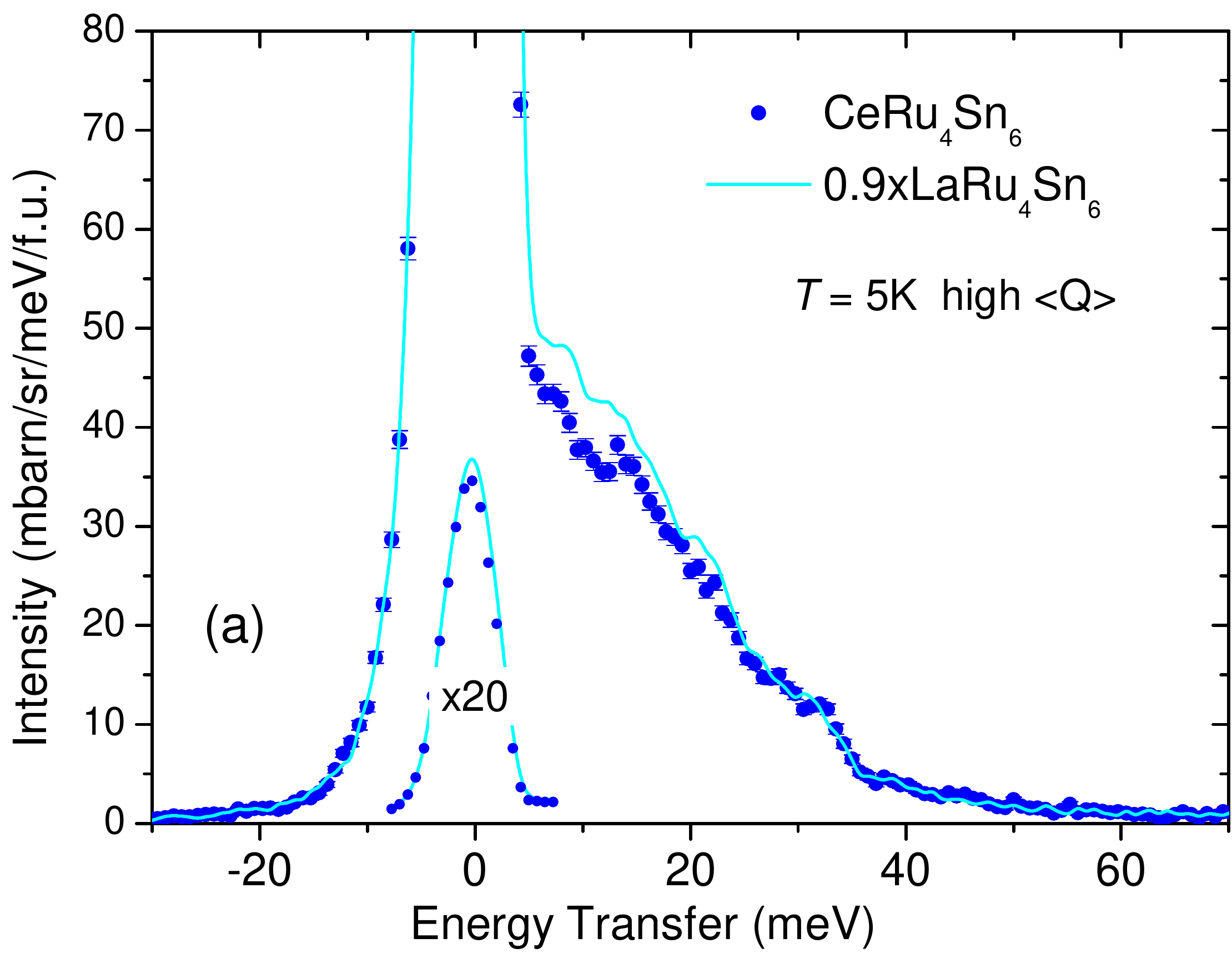}
\includegraphics[width=0.9\columnwidth]{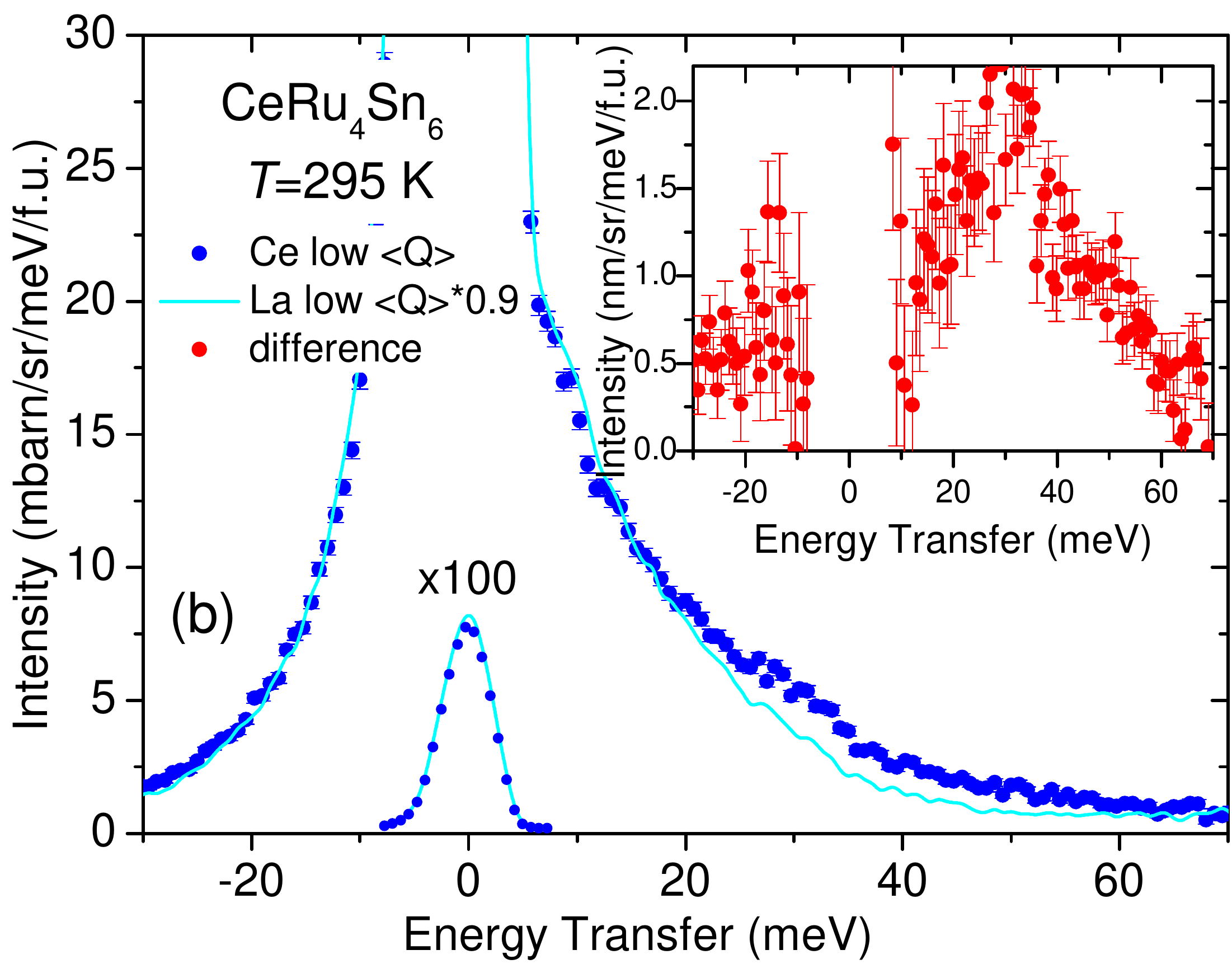}
\justify 
\label{Fig:INS}
 FIG.\,S3\, INS data with 80\,meV incident energy. (a) High angle ($<$Q$>$\,=\,9.98\,\AA$^{-1}$) neutron energy loss spectra of CeRu$_4$Sn$_6$ (blue dots) and LaRu$_4$Sn$_6$ (cyan line) data at low temperatures. The elastic scattering has been divided by a factor of 20. (b) Low angle ($<$Q$>$\,=\,2.2\,\AA$^{-1}$) CeRu$_4$Sn$_6$ (blue dots) and LaRu$_4$Sn$_6$ (cyan line) data at room temperature. The La data have been multiplied with a factor of 0.9. The elastic scattering has been divided by a factor of 100.
 \end{figure*}

\subsection{Crystal-electric field wave functions}
Table\,S1 lists the crystal-electric field wave functions for the seven Kramers doublets of the $J$\,=5/2 and 7/2 multiplets. All states are pure $J_z$ states despite the tetragonal point symmetry of Ce in CeRu$_4$Sn$_6$ because the absence of direction dependence shows that $\check A_4^4$\,=\,0 and also $\check A_6^4$\,=\,0 must be zero. The table shows the intermixing of both multiplets and how this intermixing increases for non  zero $\check A_6^0$. Note, for $\check A_6^0$\,=\,200\,meV the order of states is modified;  $|7/2,\pm 1/2 \rangle$ and $|7/2,\pm 3/2 \rangle$ exchange position.

\begin{table*}
\renewcommand{\arraystretch}{1.25}
	\begin{tabular*}{0.98\textwidth}{@{\extracolsep{\fill}} rc|rc}
		\hline
$\Delta$E$_i$  & $a|J$\,=\,5/2,\,$\pm$$J_z$$\rangle$\,+\,$b|J$\,=\,7/2,\,$\pm$$J_z$$\rangle$   &$\Delta$E$_i$  &$a|J$\,=\,5/2,\,$\pm$$J_z$$\rangle$\,+\,$b|J$\,=\,7/2,\,$\pm$$J_z$$\rangle$  \\
		     (meV)  & $\check A_2^0$\,=\,-163.7 $\check A_4^0$\,=\,-13.6 $\check A_4^4$\,=\,0 $\check A_6^0$\,=\,0 $\zeta_{LS}$\,=\,77.4\,(meV)&	  (meV)  & $\check A_2^0$\,=\,-153.5 $\check A_4^0$\,=\,-26.5 $\check A_4^4$\,=\,0 $\check A_6^0$\,=\,200 $\zeta_{LS}$\,=\,75.5(meV)\\
		\hline
		0          &       0.999$|5/2,\pm 1/2 \rangle$ + 0.024$|7/2,\pm 1/2 \rangle$ &0          &       0.993$|5/2,\pm 1/2 \rangle$ + 0.118$|7/2,\pm 1/2 \rangle$     \\
		31         &       0.998$|5/2,\pm 3/2 \rangle$ + 0.063$|7/2,\pm 3/2 \rangle$ &31         &       0.986$|5/2,\pm 3/2 \rangle$ + 0.165$|7/2,\pm 3/2 \rangle$     \\
		84         &       0.997$|5/2,\pm 5/2 \rangle$ + 0.075$|7/2,\pm 5/2 \rangle$ &84         &       0.999$|5/2,\pm 5/2 \rangle$ + 0.038$|7/2,\pm 5/2 \rangle$     \\[2mm]	
			
		270        &       0.024$|5/2,\pm 1/2 \rangle$ + 0.999$|7/2,\pm 1/2 \rangle$ &270        &       0.165$|5/2,\pm 3/2 \rangle$ + 0.986$|7/2,\pm 3/2 \rangle$     \\
		288        &       0.063$|5/2,\pm 3/2 \rangle$ + 0.998$|7/2,\pm 3/2 \rangle$ &282        &       0.118$|5/2,\pm 1/2 \rangle$ + 0.993$|7/2,\pm 1/2 \rangle$     \\
		321        &       0.075$|5/2,\pm 5/2 \rangle$ + 0.997$|7/2,\pm 5/2 \rangle$ &330        &       0.038$|5/2,\pm 5/2 \rangle$ + 0.999$|7/2,\pm 5/2 \rangle$     \\
		363        &                                     1.000$|7/2,\pm 7/2 \rangle$ &352        &                                     1.000$|7/2,\pm 7/2 \rangle$     \\
		\hline
	\end{tabular*}
\justify
 TABLE\,S1\,
Energies and wave functions of the $^2$F$_\frac{5}{2}$ and $^2$F$_\frac{7}{2}$ Kramers doublets for $\check A_6^0$\,=\,0 and 200\,meV. $a$ and $b$ denote the multiple intermixing.
\label{CF_wave_function}
\end{table*}


\begin{thebibliography}{37}%
\makeatletter
\providecommand \@ifxundefined [1]{%
 \@ifx{#1\undefined}
}%
\providecommand \@ifnum [1]{%
 \ifnum #1\expandafter \@firstoftwo
 \else \expandafter \@secondoftwo
 \fi
}%
\providecommand \@ifx [1]{%
 \ifx #1\expandafter \@firstoftwo
 \else \expandafter \@secondoftwo
 \fi
}%
\providecommand \natexlab [1]{#1}%
\providecommand \enquote  [1]{``#1''}%
\providecommand \bibnamefont  [1]{#1}%
\providecommand \bibfnamefont [1]{#1}%
\providecommand \citenamefont [1]{#1}%
\providecommand \href@noop [0]{\@secondoftwo}%
\providecommand \href [0]{\begingroup \@sanitize@url \@href}%
\providecommand \@href[1]{\@@startlink{#1}\@@href}%
\providecommand \@@href[1]{\endgroup#1\@@endlink}%
\providecommand \@sanitize@url [0]{\catcode `\\12\catcode `\$12\catcode
  `\&12\catcode `\#12\catcode `\^12\catcode `\_12\catcode `\%12\relax}%
\providecommand \@@startlink[1]{}%
\providecommand \@@endlink[0]{}%
\providecommand \url  [0]{\begingroup\@sanitize@url \@url }%
\providecommand \@url [1]{\endgroup\@href {#1}{\urlprefix }}%
\providecommand \urlprefix  [0]{URL }%
\providecommand \Eprint [0]{\href }%
\providecommand \doibase [0]{http://dx.doi.org/}%
\providecommand \selectlanguage [0]{\@gobble}%
\providecommand \bibinfo  [0]{\@secondoftwo}%
\providecommand \bibfield  [0]{\@secondoftwo}%
\providecommand \translation [1]{[#1]}%
\providecommand \BibitemOpen [0]{}%
\providecommand \bibitemStop [0]{}%
\providecommand \bibitemNoStop [0]{.\EOS\space}%
\providecommand \EOS [0]{\spacefactor3000\relax}%
\providecommand \BibitemShut  [1]{\csname bibitem#1\endcsname}%
\let\auto@bib@innerbib\@empty
\bibitem [{\citenamefont {Riseborough}(2000)}]{Riseborough2000}%
  \BibitemOpen
  \bibfield  {author} {\bibinfo {author} {\bibfnamefont {P.~S.}\ \bibnamefont
  {Riseborough}},\ }\href {\doibase 10.1080/000187300243345} {\bibfield
  {journal} {\bibinfo  {journal} {Adv. Phys.}\ }\textbf {\bibinfo {volume}
  {49(3)}},\ \bibinfo {pages} {257} (\bibinfo {year} {2000})}\BibitemShut
  {NoStop}%
\bibitem [{\citenamefont {Dzero}\ \emph {et~al.}(2010)\citenamefont {Dzero},
  \citenamefont {Sun}, \citenamefont {Galitski},\ and\ \citenamefont
  {Coleman}}]{Dzero2010}%
  \BibitemOpen
  \bibfield  {author} {\bibinfo {author} {\bibfnamefont {M.}~\bibnamefont
  {Dzero}}, \bibinfo {author} {\bibfnamefont {K.}~\bibnamefont {Sun}}, \bibinfo
  {author} {\bibfnamefont {V.}~\bibnamefont {Galitski}}, \ and\ \bibinfo
  {author} {\bibfnamefont {P.}~\bibnamefont {Coleman}},\ }\href {\doibase
  10.1103/PhysRevLett.104.106408} {\bibfield  {journal} {\bibinfo  {journal}
  {Phys. Rev. Lett.}\ }\textbf {\bibinfo {volume} {104}},\ \bibinfo {pages}
  {106408} (\bibinfo {year} {2010})}\BibitemShut {NoStop}%
\bibitem [{\citenamefont {Takimoto}(2011)}]{Takimoto2011}%
  \BibitemOpen
  \bibfield  {author} {\bibinfo {author} {\bibfnamefont {T.}~\bibnamefont
  {Takimoto}},\ }\href {\doibase 10.1143/JPSJ.80.123710} {\bibfield  {journal}
  {\bibinfo  {journal} {J. Phys. Soc. Jpn.}\ }\textbf {\bibinfo {volume}
  {80}},\ \bibinfo {pages} {123710} (\bibinfo {year} {2011})}\BibitemShut
  {NoStop}%
\bibitem [{\citenamefont {Dzero}\ \emph {et~al.}(2016)\citenamefont {Dzero},
  \citenamefont {Xia}, \citenamefont {Galitski},\ and\ \citenamefont
  {Coleman}}]{Dzero2016}%
  \BibitemOpen
  \bibfield  {author} {\bibinfo {author} {\bibfnamefont {M.}~\bibnamefont
  {Dzero}}, \bibinfo {author} {\bibfnamefont {J.}~\bibnamefont {Xia}}, \bibinfo
  {author} {\bibfnamefont {V.}~\bibnamefont {Galitski}}, \ and\ \bibinfo
  {author} {\bibfnamefont {P.}~\bibnamefont {Coleman}},\ }\href@noop {}
  {\bibfield  {journal} {\bibinfo  {journal} {Annual Rev. Con. Mat Phys.}\
  }\textbf {\bibinfo {volume} {7}} (\bibinfo {year} {2016})}\BibitemShut
  {NoStop}%
\bibitem [{\citenamefont {Dzsaber}\ \emph {et~al.}(2017)\citenamefont
  {Dzsaber}, \citenamefont {Prochaska}, \citenamefont {Sidorenko},
  \citenamefont {Eguchi}, \citenamefont {Svagera}, \citenamefont {Waas},
  \citenamefont {Prokofiev}, \citenamefont {Si},\ and\ \citenamefont
  {Paschen}}]{Dzsaber2017}%
  \BibitemOpen
  \bibfield  {author} {\bibinfo {author} {\bibfnamefont {S.}~\bibnamefont
  {Dzsaber}}, \bibinfo {author} {\bibfnamefont {L.}~\bibnamefont {Prochaska}},
  \bibinfo {author} {\bibfnamefont {A.}~\bibnamefont {Sidorenko}}, \bibinfo
  {author} {\bibfnamefont {G.}~\bibnamefont {Eguchi}}, \bibinfo {author}
  {\bibfnamefont {R.}~\bibnamefont {Svagera}}, \bibinfo {author} {\bibfnamefont
  {M.}~\bibnamefont {Waas}}, \bibinfo {author} {\bibfnamefont {A.}~\bibnamefont
  {Prokofiev}}, \bibinfo {author} {\bibfnamefont {Q.}~\bibnamefont {Si}}, \
  and\ \bibinfo {author} {\bibfnamefont {S.}~\bibnamefont {Paschen}},\ }\href
  {\doibase 10.1103/PhysRevLett.118.246601} {\bibfield  {journal} {\bibinfo
  {journal} {Phys. Rev. Lett.}\ }\textbf {\bibinfo {volume} {118}},\ \bibinfo
  {pages} {246601} (\bibinfo {year} {2017})}\BibitemShut {NoStop}%
\bibitem [{\citenamefont {Lai}\ \emph {et~al.}(2018)\citenamefont {Lai},
  \citenamefont {Grefe}, \citenamefont {Paschen},\ and\ \citenamefont
  {Si}}]{Lai93}%
  \BibitemOpen
  \bibfield  {author} {\bibinfo {author} {\bibfnamefont {H.-H.}\ \bibnamefont
  {Lai}}, \bibinfo {author} {\bibfnamefont {S.~E.}\ \bibnamefont {Grefe}},
  \bibinfo {author} {\bibfnamefont {S.}~\bibnamefont {Paschen}}, \ and\
  \bibinfo {author} {\bibfnamefont {Q.}~\bibnamefont {Si}},\ }\href {\doibase
  10.1073/pnas.1715851115} {\bibfield  {journal} {\bibinfo  {journal} {Proc.
  Nat. Acad. Sci.}\ }\textbf {\bibinfo {volume} {115}},\ \bibinfo {pages} {93}
  (\bibinfo {year} {2018})},\ \Eprint
  {http://arxiv.org/abs/http://www.pnas.org/content/115/1/93.full.pdf}
  {http://www.pnas.org/content/115/1/93.full.pdf} \BibitemShut {NoStop}%
\bibitem [{\citenamefont {P\"ottgen}\ \emph {et~al.}(1997)\citenamefont
  {P\"ottgen}, \citenamefont {Hoffman}, \citenamefont {Sampathkumaran},
  \citenamefont {Das}, \citenamefont {Mosel},\ and\ \citenamefont
  {M\"ullmann}}]{Pottgen_1997}%
  \BibitemOpen
  \bibfield  {author} {\bibinfo {author} {\bibfnamefont {R.}~\bibnamefont
  {P\"ottgen}}, \bibinfo {author} {\bibfnamefont {R.-D.}\ \bibnamefont
  {Hoffman}}, \bibinfo {author} {\bibfnamefont {E.~V.}\ \bibnamefont
  {Sampathkumaran}}, \bibinfo {author} {\bibfnamefont {I.}~\bibnamefont {Das}},
  \bibinfo {author} {\bibfnamefont {B.~D.}\ \bibnamefont {Mosel}}, \ and\
  \bibinfo {author} {\bibfnamefont {R.}~\bibnamefont {M\"ullmann}},\ }\href
  {\doibase 10.1006/jssc.1997.7565} {\bibfield  {journal} {\bibinfo  {journal}
  {J. Solid State Chem.}\ }\textbf {\bibinfo {volume} {134}},\ \bibinfo {pages}
  {326} (\bibinfo {year} {1997})}\BibitemShut {NoStop}%
\bibitem [{\citenamefont {Das}\ and\ \citenamefont
  {Sampathkumaran}(1992)}]{Das_1992}%
  \BibitemOpen
  \bibfield  {author} {\bibinfo {author} {\bibfnamefont {I.}~\bibnamefont
  {Das}}\ and\ \bibinfo {author} {\bibfnamefont {E.~V.}\ \bibnamefont
  {Sampathkumaran}},\ }\href {\doibase 10.1103/PhysRevB.46.4250} {\bibfield
  {journal} {\bibinfo  {journal} {Phys. Rev. B}\ }\textbf {\bibinfo {volume}
  {46(7)}},\ \bibinfo {pages} {4250} (\bibinfo {year} {1992})}\BibitemShut
  {NoStop}%
\bibitem [{\citenamefont {Strydom}\ \emph {et~al.}(2005)\citenamefont
  {Strydom}, \citenamefont {Guo}, \citenamefont {Paschen}, \citenamefont
  {Viennois},\ and\ \citenamefont {Steglich}}]{Strydom_2005}%
  \BibitemOpen
  \bibfield  {author} {\bibinfo {author} {\bibfnamefont {A.~M.}\ \bibnamefont
  {Strydom}}, \bibinfo {author} {\bibfnamefont {Z.}~\bibnamefont {Guo}},
  \bibinfo {author} {\bibfnamefont {S.}~\bibnamefont {Paschen}}, \bibinfo
  {author} {\bibfnamefont {R.}~\bibnamefont {Viennois}}, \ and\ \bibinfo
  {author} {\bibfnamefont {F.}~\bibnamefont {Steglich}},\ }\href {\doibase
  10.1016/j.physb.2005.01.111} {\bibfield  {journal} {\bibinfo  {journal}
  {Physica B}\ }\textbf {\bibinfo {volume} {359-361}},\ \bibinfo {pages} {293}
  (\bibinfo {year} {2005})}\BibitemShut {NoStop}%
\bibitem [{\citenamefont {Paschen}\ \emph {et~al.}(2010)\citenamefont
  {Paschen}, \citenamefont {Winkler}, \citenamefont {Nezu}, \citenamefont
  {Kriegisch}, \citenamefont {Hilscher}, \citenamefont {Custers},\ and\
  \citenamefont {Prokofiev}}]{Paschen_2010}%
  \BibitemOpen
  \bibfield  {author} {\bibinfo {author} {\bibfnamefont {S.}~\bibnamefont
  {Paschen}}, \bibinfo {author} {\bibfnamefont {H.}~\bibnamefont {Winkler}},
  \bibinfo {author} {\bibfnamefont {T.}~\bibnamefont {Nezu}}, \bibinfo {author}
  {\bibfnamefont {M.}~\bibnamefont {Kriegisch}}, \bibinfo {author}
  {\bibfnamefont {G.}~\bibnamefont {Hilscher}}, \bibinfo {author}
  {\bibfnamefont {J.}~\bibnamefont {Custers}}, \ and\ \bibinfo {author}
  {\bibfnamefont {A.}~\bibnamefont {Prokofiev}},\ }\href {\doibase
  10.1088/1742-6596/200/1/012156} {\bibfield  {journal} {\bibinfo  {journal}
  {J. Phys.: Conf. Ser.}\ }\textbf {\bibinfo {volume} {200}},\ \bibinfo {pages}
  {012156} (\bibinfo {year} {2010})}\BibitemShut {NoStop}%
\bibitem [{\citenamefont {Winkler}\ \emph {et~al.}(2012)\citenamefont
  {Winkler}, \citenamefont {Lorenzer}, \citenamefont {Prokofiev},\ and\
  \citenamefont {Paschen}}]{Winkler_2012}%
  \BibitemOpen
  \bibfield  {author} {\bibinfo {author} {\bibfnamefont {H.}~\bibnamefont
  {Winkler}}, \bibinfo {author} {\bibfnamefont {K.-A.}\ \bibnamefont
  {Lorenzer}}, \bibinfo {author} {\bibfnamefont {A.}~\bibnamefont {Prokofiev}},
  \ and\ \bibinfo {author} {\bibfnamefont {S.}~\bibnamefont {Paschen}},\ }\href
  {\doibase 10.1088/1742-6596/391/1/012077} {\bibfield  {journal} {\bibinfo
  {journal} {J. Phys.: Conf. Ser.}\ }\textbf {\bibinfo {volume} {391}},\
  \bibinfo {pages} {012077} (\bibinfo {year} {2012})}\BibitemShut {NoStop}%
\bibitem [{\citenamefont {Guritanu}\ \emph {et~al.}(2013)\citenamefont
  {Guritanu}, \citenamefont {Wissgott}, \citenamefont {Weig}, \citenamefont
  {Winkler}, \citenamefont {Sichelschmidt}, \citenamefont {Scheffler},
  \citenamefont {Prokofiev}, \citenamefont {Kimura}, \citenamefont {Iizuka},
  \citenamefont {Strydom}, \citenamefont {Dressel}, \citenamefont {Steglich},
  \citenamefont {Held},\ and\ \citenamefont {Paschen}}]{Guritanu_2013}%
  \BibitemOpen
  \bibfield  {author} {\bibinfo {author} {\bibfnamefont {V.}~\bibnamefont
  {Guritanu}}, \bibinfo {author} {\bibfnamefont {P.}~\bibnamefont {Wissgott}},
  \bibinfo {author} {\bibfnamefont {T.}~\bibnamefont {Weig}}, \bibinfo {author}
  {\bibfnamefont {H.}~\bibnamefont {Winkler}}, \bibinfo {author} {\bibfnamefont
  {J.}~\bibnamefont {Sichelschmidt}}, \bibinfo {author} {\bibfnamefont
  {M.}~\bibnamefont {Scheffler}}, \bibinfo {author} {\bibfnamefont
  {A.}~\bibnamefont {Prokofiev}}, \bibinfo {author} {\bibfnamefont
  {S.}~\bibnamefont {Kimura}}, \bibinfo {author} {\bibfnamefont
  {T.}~\bibnamefont {Iizuka}}, \bibinfo {author} {\bibfnamefont {A.~M.}\
  \bibnamefont {Strydom}}, \bibinfo {author} {\bibfnamefont {M.}~\bibnamefont
  {Dressel}}, \bibinfo {author} {\bibfnamefont {F.}~\bibnamefont {Steglich}},
  \bibinfo {author} {\bibfnamefont {K.}~\bibnamefont {Held}}, \ and\ \bibinfo
  {author} {\bibfnamefont {S.}~\bibnamefont {Paschen}},\ }\href {\doibase
  10.1103/PhysRevB.87.115129} {\bibfield  {journal} {\bibinfo  {journal} {Phys.
  Rev. B}\ }\textbf {\bibinfo {volume} {87(11)}},\ \bibinfo {pages} {115129}
  (\bibinfo {year} {2013})}\BibitemShut {NoStop}%
\bibitem [{\citenamefont {Strydom}\ \emph {et~al.}(2007)\citenamefont
  {Strydom}, \citenamefont {Hillier}, \citenamefont {Adroja}, \citenamefont
  {Paschen},\ and\ \citenamefont {Steglich}}]{Strydom_2007}%
  \BibitemOpen
  \bibfield  {author} {\bibinfo {author} {\bibfnamefont {A.~M.}\ \bibnamefont
  {Strydom}}, \bibinfo {author} {\bibfnamefont {A.~D.}\ \bibnamefont
  {Hillier}}, \bibinfo {author} {\bibfnamefont {D.~T.}\ \bibnamefont {Adroja}},
  \bibinfo {author} {\bibfnamefont {S.}~\bibnamefont {Paschen}}, \ and\
  \bibinfo {author} {\bibfnamefont {F.}~\bibnamefont {Steglich}},\ }\href
  {\doibase 10.1016/j.jmmm.2006.10.084} {\bibfield  {journal} {\bibinfo
  {journal} {J. Magn. Magn. Mater.}\ }\textbf {\bibinfo {volume} {310}},\
  \bibinfo {pages} {377} (\bibinfo {year} {2007})}\BibitemShut {NoStop}%
\bibitem [{\citenamefont {Sundermann}\ \emph {et~al.}(2015)\citenamefont
  {Sundermann}, \citenamefont {Strigari}, \citenamefont {Willers},
  \citenamefont {Winkler}, \citenamefont {Prokofiev}, \citenamefont {Ablett},
  \citenamefont {Rueff}, \citenamefont {Schmitz}, \citenamefont {Weschke},
  \citenamefont {Sala}, \citenamefont {Al-Zein}, \citenamefont {Tanaka},
  \citenamefont {Haverkort}, \citenamefont {Kasinathan}, \citenamefont {Tjeng},
  \citenamefont {Paschen},\ and\ \citenamefont {Severing}}]{Sundermann2015}%
  \BibitemOpen
  \bibfield  {author} {\bibinfo {author} {\bibfnamefont {M.}~\bibnamefont
  {Sundermann}}, \bibinfo {author} {\bibfnamefont {F.}~\bibnamefont
  {Strigari}}, \bibinfo {author} {\bibfnamefont {T.}~\bibnamefont {Willers}},
  \bibinfo {author} {\bibfnamefont {H.}~\bibnamefont {Winkler}}, \bibinfo
  {author} {\bibfnamefont {A.}~\bibnamefont {Prokofiev}}, \bibinfo {author}
  {\bibfnamefont {J.~M.}\ \bibnamefont {Ablett}}, \bibinfo {author}
  {\bibfnamefont {J.-P.}\ \bibnamefont {Rueff}}, \bibinfo {author}
  {\bibfnamefont {D.}~\bibnamefont {Schmitz}}, \bibinfo {author} {\bibfnamefont
  {E.}~\bibnamefont {Weschke}}, \bibinfo {author} {\bibfnamefont {M.~M.}\
  \bibnamefont {Sala}}, \bibinfo {author} {\bibfnamefont {A.}~\bibnamefont
  {Al-Zein}}, \bibinfo {author} {\bibfnamefont {A.}~\bibnamefont {Tanaka}},
  \bibinfo {author} {\bibfnamefont {M.~W.}\ \bibnamefont {Haverkort}}, \bibinfo
  {author} {\bibfnamefont {D.}~\bibnamefont {Kasinathan}}, \bibinfo {author}
  {\bibfnamefont {L.~H.}\ \bibnamefont {Tjeng}}, \bibinfo {author}
  {\bibfnamefont {S.}~\bibnamefont {Paschen}}, \ and\ \bibinfo {author}
  {\bibfnamefont {A.}~\bibnamefont {Severing}},\ }\href {\doibase
  10.1038/srep17937} {\bibfield  {journal} {\bibinfo  {journal} {Sci. Rep.}\
  }\textbf {\bibinfo {volume} {5}},\ \bibinfo {pages} {17937} (\bibinfo {year}
  {2015})}\BibitemShut {NoStop}%
\bibitem [{\citenamefont {Sundermann}\ \emph {et~al.}(2017)\citenamefont
  {Sundermann}, \citenamefont {Chen}, \citenamefont {Utsumi}, \citenamefont
  {Wu}, \citenamefont {Tsuei}, \citenamefont {Haenel}, \citenamefont
  {Prokofiev}, \citenamefont {Paschen}, \citenamefont {Tanaka}, \citenamefont
  {Tjeng},\ and\ \citenamefont {Severing}}]{Sundermann2017}%
  \BibitemOpen
  \bibfield  {author} {\bibinfo {author} {\bibfnamefont {M.}~\bibnamefont
  {Sundermann}}, \bibinfo {author} {\bibfnamefont {K.}~\bibnamefont {Chen}},
  \bibinfo {author} {\bibfnamefont {Y.}~\bibnamefont {Utsumi}}, \bibinfo
  {author} {\bibfnamefont {Y.-H.}\ \bibnamefont {Wu}}, \bibinfo {author}
  {\bibfnamefont {K.-D.}\ \bibnamefont {Tsuei}}, \bibinfo {author}
  {\bibfnamefont {J.}~\bibnamefont {Haenel}}, \bibinfo {author} {\bibfnamefont
  {A.}~\bibnamefont {Prokofiev}}, \bibinfo {author} {\bibfnamefont
  {S.}~\bibnamefont {Paschen}}, \bibinfo {author} {\bibfnamefont
  {A.}~\bibnamefont {Tanaka}}, \bibinfo {author} {\bibfnamefont {L.~H.}\
  \bibnamefont {Tjeng}}, \ and\ \bibinfo {author} {\bibfnamefont
  {A.}~\bibnamefont {Severing}},\ }\href
  {http://stacks.iop.org/1742-6596/807/i=2/a=022001} {\bibfield  {journal}
  {\bibinfo  {journal} {Journal of Physics: Conference Series}\ }\textbf
  {\bibinfo {volume} {807}},\ \bibinfo {pages} {022001} (\bibinfo {year}
  {2017})}\BibitemShut {NoStop}%
\bibitem [{\citenamefont {Xu}\ \emph {et~al.}(2017)\citenamefont {Xu},
  \citenamefont {Yue}, \citenamefont {Weng},\ and\ \citenamefont
  {Dai}}]{Xu2017}%
  \BibitemOpen
  \bibfield  {author} {\bibinfo {author} {\bibfnamefont {Y.}~\bibnamefont
  {Xu}}, \bibinfo {author} {\bibfnamefont {C.}~\bibnamefont {Yue}}, \bibinfo
  {author} {\bibfnamefont {H.}~\bibnamefont {Weng}}, \ and\ \bibinfo {author}
  {\bibfnamefont {X.}~\bibnamefont {Dai}},\ }\href {\doibase
  10.1103/PhysRevX.7.011027} {\bibfield  {journal} {\bibinfo  {journal} {Phys.
  Rev. X}\ }\textbf {\bibinfo {volume} {7}},\ \bibinfo {pages} {011027}
  (\bibinfo {year} {2017})}\BibitemShut {NoStop}%
\bibitem [{\citenamefont {Amorese}\ \emph {et~al.}(2016)\citenamefont
  {Amorese}, \citenamefont {Dellea}, \citenamefont {Fanciulli}, \citenamefont
  {Seiro}, \citenamefont {Geibel}, \citenamefont {Krellner}, \citenamefont
  {Makarova}, \citenamefont {Braicovich}, \citenamefont {Ghiringhelli},
  \citenamefont {Vyalikh}, \citenamefont {Brookes},\ and\ \citenamefont
  {Kummer}}]{Amorese_2016}%
  \BibitemOpen
  \bibfield  {author} {\bibinfo {author} {\bibfnamefont {A.}~\bibnamefont
  {Amorese}}, \bibinfo {author} {\bibfnamefont {G.}~\bibnamefont {Dellea}},
  \bibinfo {author} {\bibfnamefont {M.}~\bibnamefont {Fanciulli}}, \bibinfo
  {author} {\bibfnamefont {S.}~\bibnamefont {Seiro}}, \bibinfo {author}
  {\bibfnamefont {C.}~\bibnamefont {Geibel}}, \bibinfo {author} {\bibfnamefont
  {C.}~\bibnamefont {Krellner}}, \bibinfo {author} {\bibfnamefont {I.~P.}\
  \bibnamefont {Makarova}}, \bibinfo {author} {\bibfnamefont {L.}~\bibnamefont
  {Braicovich}}, \bibinfo {author} {\bibfnamefont {G.}~\bibnamefont
  {Ghiringhelli}}, \bibinfo {author} {\bibfnamefont {D.~V.}\ \bibnamefont
  {Vyalikh}}, \bibinfo {author} {\bibfnamefont {N.~B.}\ \bibnamefont
  {Brookes}}, \ and\ \bibinfo {author} {\bibfnamefont {K.}~\bibnamefont
  {Kummer}},\ }\href {\doibase 10.1103/PhysRevB.93.165134} {\bibfield
  {journal} {\bibinfo  {journal} {Phys. Rev. B}\ }\textbf {\bibinfo {volume}
  {93}},\ \bibinfo {pages} {165134} (\bibinfo {year} {2016})}\BibitemShut
  {NoStop}%
\bibitem [{\citenamefont {Amorese}\ \emph
  {et~al.}(2018{\natexlab{a}})\citenamefont {Amorese}, \citenamefont
  {Caroca-Canales}, \citenamefont {Seiro}, \citenamefont {Krellner},
  \citenamefont {Ghiringhelli}, \citenamefont {Brookes}, \citenamefont
  {Vyalikh},\ and\ \citenamefont {Kummer}}]{Amorese_2018}%
  \BibitemOpen
  \bibfield  {author} {\bibinfo {author} {\bibfnamefont {A.}~\bibnamefont
  {Amorese}}, \bibinfo {author} {\bibfnamefont {N.}~\bibnamefont
  {Caroca-Canales}}, \bibinfo {author} {\bibfnamefont {S.}~\bibnamefont
  {Seiro}}, \bibinfo {author} {\bibfnamefont {C.}~\bibnamefont {Krellner}},
  \bibinfo {author} {\bibfnamefont {G.}~\bibnamefont {Ghiringhelli}}, \bibinfo
  {author} {\bibfnamefont {N.~B.}\ \bibnamefont {Brookes}}, \bibinfo {author}
  {\bibfnamefont {C.}~\bibnamefont {Vyalikh}, \bibfnamefont {D.~V.and~Geibel}},
  \ and\ \bibinfo {author} {\bibfnamefont {K.}~\bibnamefont {Kummer}},\
  }\href@noop {} {} (\bibinfo {year} {2018}{\natexlab{a}}),\ \bibinfo {note}
  {submitted and arXiv:1803.11068}\BibitemShut {NoStop}%
\bibitem [{\citenamefont {Nakazawa}\ \emph {et~al.}(1996)\citenamefont
  {Nakazawa}, \citenamefont {Tanaka}, \citenamefont {Uozumi},\ and\
  \citenamefont {Kotani}}]{Nakazawa_1996}%
  \BibitemOpen
  \bibfield  {author} {\bibinfo {author} {\bibfnamefont {M.}~\bibnamefont
  {Nakazawa}}, \bibinfo {author} {\bibfnamefont {S.}~\bibnamefont {Tanaka}},
  \bibinfo {author} {\bibfnamefont {T.}~\bibnamefont {Uozumi}}, \ and\ \bibinfo
  {author} {\bibfnamefont {A.}~\bibnamefont {Kotani}},\ }\href {\doibase
  10.1143/JPSJ.65.2303} {\bibfield  {journal} {\bibinfo  {journal} {Journal of
  the Physical Society of Japan}\ }\textbf {\bibinfo {volume} {65}},\ \bibinfo
  {pages} {2303} (\bibinfo {year} {1996})},\ \Eprint
  {http://arxiv.org/abs/https://doi.org/10.1143/JPSJ.65.2303}
  {https://doi.org/10.1143/JPSJ.65.2303} \BibitemShut {NoStop}%
\bibitem [{\citenamefont {Butorin}\ \emph {et~al.}(1996)\citenamefont
  {Butorin}, \citenamefont {Mancini}, \citenamefont {Guo}, \citenamefont
  {Wassdahl}, \citenamefont {Nordgren}, \citenamefont {Nakazawa}, \citenamefont
  {Tanaka}, \citenamefont {Uozumi}, \citenamefont {Kotani}, \citenamefont {Ma},
  \citenamefont {Myano}, \citenamefont {Karlin},\ and\ \citenamefont
  {Shuh}}]{Butorin_1996}%
  \BibitemOpen
  \bibfield  {author} {\bibinfo {author} {\bibfnamefont {S.~M.}\ \bibnamefont
  {Butorin}}, \bibinfo {author} {\bibfnamefont {D.~C.}\ \bibnamefont
  {Mancini}}, \bibinfo {author} {\bibfnamefont {J.-H.}\ \bibnamefont {Guo}},
  \bibinfo {author} {\bibfnamefont {N.}~\bibnamefont {Wassdahl}}, \bibinfo
  {author} {\bibfnamefont {J.}~\bibnamefont {Nordgren}}, \bibinfo {author}
  {\bibfnamefont {M.}~\bibnamefont {Nakazawa}}, \bibinfo {author}
  {\bibfnamefont {S.}~\bibnamefont {Tanaka}}, \bibinfo {author} {\bibfnamefont
  {T.}~\bibnamefont {Uozumi}}, \bibinfo {author} {\bibfnamefont
  {A.}~\bibnamefont {Kotani}}, \bibinfo {author} {\bibfnamefont
  {Y.}~\bibnamefont {Ma}}, \bibinfo {author} {\bibfnamefont {K.~E.}\
  \bibnamefont {Myano}}, \bibinfo {author} {\bibfnamefont {B.~A.}\ \bibnamefont
  {Karlin}}, \ and\ \bibinfo {author} {\bibfnamefont {D.~K.}\ \bibnamefont
  {Shuh}},\ }\href {\doibase 10.1103/PhysRevLett.77.574} {\bibfield  {journal}
  {\bibinfo  {journal} {Phys. Rev. Lett.}\ }\textbf {\bibinfo {volume} {77}},\
  \bibinfo {pages} {574} (\bibinfo {year} {1996})}\BibitemShut {NoStop}%
\bibitem [{\citenamefont {Dallera}\ \emph {et~al.}(2001)\citenamefont
  {Dallera}, \citenamefont {Giarda}, \citenamefont {Ghiringhelli},
  \citenamefont {Tagliaferri}, \citenamefont {Braicovich},\ and\ \citenamefont
  {Brookes}}]{Dallera_2001}%
  \BibitemOpen
  \bibfield  {author} {\bibinfo {author} {\bibfnamefont {C.}~\bibnamefont
  {Dallera}}, \bibinfo {author} {\bibfnamefont {K.}~\bibnamefont {Giarda}},
  \bibinfo {author} {\bibfnamefont {G.}~\bibnamefont {Ghiringhelli}}, \bibinfo
  {author} {\bibfnamefont {A.}~\bibnamefont {Tagliaferri}}, \bibinfo {author}
  {\bibfnamefont {L.}~\bibnamefont {Braicovich}}, \ and\ \bibinfo {author}
  {\bibfnamefont {N.~B.}\ \bibnamefont {Brookes}},\ }\href {\doibase
  10.1103/PhysRevB.64.153104} {\bibfield  {journal} {\bibinfo  {journal} {Phys.
  Rev. B}\ }\textbf {\bibinfo {volume} {64}},\ \bibinfo {pages} {153104}
  (\bibinfo {year} {2001})}\BibitemShut {NoStop}%
\bibitem [{\citenamefont {Ament}\ \emph {et~al.}(2011)\citenamefont {Ament},
  \citenamefont {van Veenendaal},\ and\ \citenamefont {van~den
  Brink}}]{Ament_2011b}%
  \BibitemOpen
  \bibfield  {author} {\bibinfo {author} {\bibfnamefont {L.~J.~P.}\
  \bibnamefont {Ament}}, \bibinfo {author} {\bibfnamefont {M.}~\bibnamefont
  {van Veenendaal}}, \ and\ \bibinfo {author} {\bibfnamefont {J.}~\bibnamefont
  {van~den Brink}},\ }\href {http://stacks.iop.org/0295-5075/95/i=2/a=27008}
  {\bibfield  {journal} {\bibinfo  {journal} {EPL (Europhysics Letters)}\
  }\textbf {\bibinfo {volume} {95}},\ \bibinfo {pages} {27008} (\bibinfo {year}
  {2011})}\BibitemShut {NoStop}%
\bibitem [{\citenamefont {Braicovich}\ \emph {et~al.}(2010)\citenamefont
  {Braicovich}, \citenamefont {Moretti~Sala}, \citenamefont {Ament},
  \citenamefont {Bisogni}, \citenamefont {Minola}, \citenamefont {Balestrino},
  \citenamefont {Di~Castro}, \citenamefont {De~Luca}, \citenamefont {Salluzzo},
  \citenamefont {Ghiringhelli},\ and\ \citenamefont {van~den
  Brink}}]{Braicovich_2010b}%
  \BibitemOpen
  \bibfield  {author} {\bibinfo {author} {\bibfnamefont {L.}~\bibnamefont
  {Braicovich}}, \bibinfo {author} {\bibfnamefont {M.}~\bibnamefont
  {Moretti~Sala}}, \bibinfo {author} {\bibfnamefont {L.~J.~P.}\ \bibnamefont
  {Ament}}, \bibinfo {author} {\bibfnamefont {V.}~\bibnamefont {Bisogni}},
  \bibinfo {author} {\bibfnamefont {M.}~\bibnamefont {Minola}}, \bibinfo
  {author} {\bibfnamefont {G.}~\bibnamefont {Balestrino}}, \bibinfo {author}
  {\bibfnamefont {D.}~\bibnamefont {Di~Castro}}, \bibinfo {author}
  {\bibfnamefont {G.~M.}\ \bibnamefont {De~Luca}}, \bibinfo {author}
  {\bibfnamefont {M.}~\bibnamefont {Salluzzo}}, \bibinfo {author}
  {\bibfnamefont {G.}~\bibnamefont {Ghiringhelli}}, \ and\ \bibinfo {author}
  {\bibfnamefont {J.}~\bibnamefont {van~den Brink}},\ }\href {\doibase
  10.1103/PhysRevB.81.174533} {\bibfield  {journal} {\bibinfo  {journal} {Phys.
  Rev. B}\ }\textbf {\bibinfo {volume} {81}},\ \bibinfo {pages} {174533}
  (\bibinfo {year} {2010})}\BibitemShut {NoStop}%
\bibitem [{\citenamefont {Sala}\ \emph {et~al.}(2011)\citenamefont {Sala},
  \citenamefont {Bisogni}, \citenamefont {Aruta}, \citenamefont {Balestrino},
  \citenamefont {Berger}, \citenamefont {Brookes}, \citenamefont {de~Luca},
  \citenamefont {Castro}, \citenamefont {Grioni}, \citenamefont {Guarise},
  \citenamefont {Medaglia}, \citenamefont {Granozio}, \citenamefont {Minola},
  \citenamefont {Perna}, \citenamefont {Radovic}, \citenamefont {Salluzzo},
  \citenamefont {Schmitt}, \citenamefont {Zhou}, \citenamefont {Braicovich},\
  and\ \citenamefont {Ghiringhelli}}]{Moretti_2011}%
  \BibitemOpen
  \bibfield  {author} {\bibinfo {author} {\bibfnamefont {M.~M.}\ \bibnamefont
  {Sala}}, \bibinfo {author} {\bibfnamefont {V.}~\bibnamefont {Bisogni}},
  \bibinfo {author} {\bibfnamefont {C.}~\bibnamefont {Aruta}}, \bibinfo
  {author} {\bibfnamefont {G.}~\bibnamefont {Balestrino}}, \bibinfo {author}
  {\bibfnamefont {H.}~\bibnamefont {Berger}}, \bibinfo {author} {\bibfnamefont
  {N.~B.}\ \bibnamefont {Brookes}}, \bibinfo {author} {\bibfnamefont {G.~M.}\
  \bibnamefont {de~Luca}}, \bibinfo {author} {\bibfnamefont {D.~D.}\
  \bibnamefont {Castro}}, \bibinfo {author} {\bibfnamefont {M.}~\bibnamefont
  {Grioni}}, \bibinfo {author} {\bibfnamefont {M.}~\bibnamefont {Guarise}},
  \bibinfo {author} {\bibfnamefont {P.~G.}\ \bibnamefont {Medaglia}}, \bibinfo
  {author} {\bibfnamefont {F.~M.}\ \bibnamefont {Granozio}}, \bibinfo {author}
  {\bibfnamefont {M.}~\bibnamefont {Minola}}, \bibinfo {author} {\bibfnamefont
  {P.}~\bibnamefont {Perna}}, \bibinfo {author} {\bibfnamefont
  {M.}~\bibnamefont {Radovic}}, \bibinfo {author} {\bibfnamefont
  {M.}~\bibnamefont {Salluzzo}}, \bibinfo {author} {\bibfnamefont
  {T.}~\bibnamefont {Schmitt}}, \bibinfo {author} {\bibfnamefont {K.~J.}\
  \bibnamefont {Zhou}}, \bibinfo {author} {\bibfnamefont {L.}~\bibnamefont
  {Braicovich}}, \ and\ \bibinfo {author} {\bibfnamefont {G.}~\bibnamefont
  {Ghiringhelli}},\ }\href {http://stacks.iop.org/1367-2630/13/i=4/a=043026}
  {\bibfield  {journal} {\bibinfo  {journal} {New Journal of Physics}\ }\textbf
  {\bibinfo {volume} {13}},\ \bibinfo {pages} {043026} (\bibinfo {year}
  {2011})}\BibitemShut {NoStop}%
\bibitem [{\citenamefont {Minola}\ \emph {et~al.}(2015)\citenamefont {Minola},
  \citenamefont {Dellea}, \citenamefont {Gretarsson}, \citenamefont {Peng},
  \citenamefont {Lu}, \citenamefont {Porras}, \citenamefont {Loew},
  \citenamefont {Yakhou}, \citenamefont {Brookes}, \citenamefont {Huang},
  \citenamefont {Pelliciari}, \citenamefont {Schmitt}, \citenamefont
  {Ghiringhelli}, \citenamefont {Keimer}, \citenamefont {Braicovich},\ and\
  \citenamefont {Le~Tacon}}]{Minola_2015}%
  \BibitemOpen
  \bibfield  {author} {\bibinfo {author} {\bibfnamefont {M.}~\bibnamefont
  {Minola}}, \bibinfo {author} {\bibfnamefont {G.}~\bibnamefont {Dellea}},
  \bibinfo {author} {\bibfnamefont {H.}~\bibnamefont {Gretarsson}}, \bibinfo
  {author} {\bibfnamefont {Y.~Y.}\ \bibnamefont {Peng}}, \bibinfo {author}
  {\bibfnamefont {Y.}~\bibnamefont {Lu}}, \bibinfo {author} {\bibfnamefont
  {J.}~\bibnamefont {Porras}}, \bibinfo {author} {\bibfnamefont
  {T.}~\bibnamefont {Loew}}, \bibinfo {author} {\bibfnamefont {F.}~\bibnamefont
  {Yakhou}}, \bibinfo {author} {\bibfnamefont {N.~B.}\ \bibnamefont {Brookes}},
  \bibinfo {author} {\bibfnamefont {Y.~B.}\ \bibnamefont {Huang}}, \bibinfo
  {author} {\bibfnamefont {J.}~\bibnamefont {Pelliciari}}, \bibinfo {author}
  {\bibfnamefont {T.}~\bibnamefont {Schmitt}}, \bibinfo {author} {\bibfnamefont
  {G.}~\bibnamefont {Ghiringhelli}}, \bibinfo {author} {\bibfnamefont
  {B.}~\bibnamefont {Keimer}}, \bibinfo {author} {\bibfnamefont
  {L.}~\bibnamefont {Braicovich}}, \ and\ \bibinfo {author} {\bibfnamefont
  {M.}~\bibnamefont {Le~Tacon}},\ }\href {\doibase
  10.1103/PhysRevLett.114.217003} {\bibfield  {journal} {\bibinfo  {journal}
  {Phys. Rev. Lett.}\ }\textbf {\bibinfo {volume} {114}},\ \bibinfo {pages}
  {217003} (\bibinfo {year} {2015})}\BibitemShut {NoStop}%
\bibitem [{\citenamefont {Braicovich}\ \emph {et~al.}(2014)\citenamefont
  {Braicovich}, \citenamefont {Minola}, \citenamefont {Dellea}, \citenamefont
  {Tacon}, \citenamefont {Sala}, \citenamefont {Morawe}, \citenamefont
  {Peffen}, \citenamefont {Supruangnet}, \citenamefont {Yakhou}, \citenamefont
  {Ghiringhelli},\ and\ \citenamefont {Brookes}}]{Braicovich_2014}%
  \BibitemOpen
  \bibfield  {author} {\bibinfo {author} {\bibfnamefont {L.}~\bibnamefont
  {Braicovich}}, \bibinfo {author} {\bibfnamefont {M.}~\bibnamefont {Minola}},
  \bibinfo {author} {\bibfnamefont {G.}~\bibnamefont {Dellea}}, \bibinfo
  {author} {\bibfnamefont {M.~L.}\ \bibnamefont {Tacon}}, \bibinfo {author}
  {\bibfnamefont {M.~M.}\ \bibnamefont {Sala}}, \bibinfo {author}
  {\bibfnamefont {C.}~\bibnamefont {Morawe}}, \bibinfo {author} {\bibfnamefont
  {J.-C.}\ \bibnamefont {Peffen}}, \bibinfo {author} {\bibfnamefont
  {R.}~\bibnamefont {Supruangnet}}, \bibinfo {author} {\bibfnamefont
  {F.}~\bibnamefont {Yakhou}}, \bibinfo {author} {\bibfnamefont
  {G.}~\bibnamefont {Ghiringhelli}}, \ and\ \bibinfo {author} {\bibfnamefont
  {N.~B.}\ \bibnamefont {Brookes}},\ }\href {\doibase 10.1063/1.4900959}
  {\bibfield  {journal} {\bibinfo  {journal} {Rev. of Scientific Instruments}\
  }\textbf {\bibinfo {volume} {85}},\ \bibinfo {pages} {115104} (\bibinfo
  {year} {2014})},\ \Eprint
  {http://arxiv.org/abs/https://doi.org/10.1063/1.4900959}
  {https://doi.org/10.1063/1.4900959} \BibitemShut {NoStop}%
\bibitem [{\citenamefont {Haverkort}(2016)}]{Haverkort2016}%
  \BibitemOpen
  \bibfield  {author} {\bibinfo {author} {\bibfnamefont {M.~W.}\ \bibnamefont
  {Haverkort}},\ }\href {\doibase 10.1088/1742-6596/712/1/012001} {\bibfield
  {journal} {\bibinfo  {journal} {J. Phys.: Conf. Ser.}\ }\textbf {\bibinfo
  {volume} {712}},\ \bibinfo {pages} {012001} (\bibinfo {year}
  {2016})}\BibitemShut {NoStop}%
\bibitem [{\citenamefont {Haverkort}\ \emph {et~al.}(2012)\citenamefont
  {Haverkort}, \citenamefont {Zwierzycki},\ and\ \citenamefont
  {Andersen}}]{Haverkort_2012}%
  \BibitemOpen
  \bibfield  {author} {\bibinfo {author} {\bibfnamefont {M.~W.}\ \bibnamefont
  {Haverkort}}, \bibinfo {author} {\bibfnamefont {M.}~\bibnamefont
  {Zwierzycki}}, \ and\ \bibinfo {author} {\bibfnamefont {O.~K.}\ \bibnamefont
  {Andersen}},\ }\href {\doibase 10.1103/PhysRevB.85.165113} {\bibfield
  {journal} {\bibinfo  {journal} {Phys. Rev. B}\ }\textbf {\bibinfo {volume}
  {85}},\ \bibinfo {pages} {165113} (\bibinfo {year} {2012})}\BibitemShut
  {NoStop}%
\bibitem [{\citenamefont {Cowan}(1981)}]{Cowan}%
  \BibitemOpen
  \bibfield  {author} {\bibinfo {author} {\bibfnamefont {R.~D.}\ \bibnamefont
  {Cowan}},\ }\href@noop {} {\emph {\bibinfo {title} {The Theory of Atomic
  Structure and Spectra}}}\ (\bibinfo  {publisher} {University of California
  Press},\ \bibinfo {address} {Berkeley},\ \bibinfo {year} {1981})\BibitemShut
  {NoStop}%
\bibitem [{\citenamefont {Hansmann}\ \emph {et~al.}(2008)\citenamefont
  {Hansmann}, \citenamefont {Severing}, \citenamefont {Hu}, \citenamefont
  {Haverkort}, \citenamefont {Chang}, \citenamefont {Klein}, \citenamefont
  {Tanaka}, \citenamefont {Hsieh}, \citenamefont {Lin}, \citenamefont {Chen},
  \citenamefont {F\aa{}k}, \citenamefont {Lejay},\ and\ \citenamefont
  {Tjeng}}]{Hansmann2008}%
  \BibitemOpen
  \bibfield  {author} {\bibinfo {author} {\bibfnamefont {P.}~\bibnamefont
  {Hansmann}}, \bibinfo {author} {\bibfnamefont {A.}~\bibnamefont {Severing}},
  \bibinfo {author} {\bibfnamefont {Z.}~\bibnamefont {Hu}}, \bibinfo {author}
  {\bibfnamefont {M.~W.}\ \bibnamefont {Haverkort}}, \bibinfo {author}
  {\bibfnamefont {C.~F.}\ \bibnamefont {Chang}}, \bibinfo {author}
  {\bibfnamefont {S.}~\bibnamefont {Klein}}, \bibinfo {author} {\bibfnamefont
  {A.}~\bibnamefont {Tanaka}}, \bibinfo {author} {\bibfnamefont {H.~H.}\
  \bibnamefont {Hsieh}}, \bibinfo {author} {\bibfnamefont {H.-J.}\ \bibnamefont
  {Lin}}, \bibinfo {author} {\bibfnamefont {C.~T.}\ \bibnamefont {Chen}},
  \bibinfo {author} {\bibfnamefont {B.}~\bibnamefont {F\aa{}k}}, \bibinfo
  {author} {\bibfnamefont {P.}~\bibnamefont {Lejay}}, \ and\ \bibinfo {author}
  {\bibfnamefont {L.~H.}\ \bibnamefont {Tjeng}},\ }\href {\doibase
  10.1103/PhysRevLett.100.066405} {\bibfield  {journal} {\bibinfo  {journal}
  {Phys. Rev. Lett.}\ }\textbf {\bibinfo {volume} {100}},\ \bibinfo {pages}
  {066405} (\bibinfo {year} {2008})}\BibitemShut {NoStop}%
\bibitem [{\citenamefont {Tanaka}\ and\ \citenamefont {Jo}(1994)}]{Tanaka1994}%
  \BibitemOpen
  \bibfield  {author} {\bibinfo {author} {\bibfnamefont {A.}~\bibnamefont
  {Tanaka}}\ and\ \bibinfo {author} {\bibfnamefont {T.}~\bibnamefont {Jo}},\
  }\href {\doibase 10.1143/JPSJ.63.2788} {\bibfield  {journal} {\bibinfo
  {journal} {J Phys. Soc. Jpn.}\ }\textbf {\bibinfo {volume} {63}},\ \bibinfo
  {pages} {2788} (\bibinfo {year} {1994})}\BibitemShut {NoStop}%
\bibitem [{\citenamefont {Amorese}\ \emph
  {et~al.}(2018{\natexlab{b}})\citenamefont {Amorese}, \citenamefont {Kummer},
  \citenamefont {Brookes}, \citenamefont {Stockert}, \citenamefont {Adroja},
  \citenamefont {Strydom}, \citenamefont {Sidorenko}, \citenamefont {Winkler},
  \citenamefont {Zocoo}, \citenamefont {Prokofiev}, \citenamefont {Paschen},
  \citenamefont {Haverkort}, \citenamefont {Tjeng},\ and\ \citenamefont
  {Severing}}]{Supp}%
  \BibitemOpen
  \bibfield  {author} {\bibinfo {author} {\bibfnamefont {A.}~\bibnamefont
  {Amorese}}, \bibinfo {author} {\bibfnamefont {K.}~\bibnamefont {Kummer}},
  \bibinfo {author} {\bibfnamefont {N.~B.}\ \bibnamefont {Brookes}}, \bibinfo
  {author} {\bibfnamefont {O.}~\bibnamefont {Stockert}}, \bibinfo {author}
  {\bibfnamefont {D.~T.}\ \bibnamefont {Adroja}}, \bibinfo {author}
  {\bibfnamefont {A.~M.}\ \bibnamefont {Strydom}}, \bibinfo {author}
  {\bibfnamefont {A.}~\bibnamefont {Sidorenko}}, \bibinfo {author}
  {\bibfnamefont {H.}~\bibnamefont {Winkler}}, \bibinfo {author} {\bibfnamefont
  {D.~A.}\ \bibnamefont {Zocoo}}, \bibinfo {author} {\bibfnamefont
  {A.}~\bibnamefont {Prokofiev}}, \bibinfo {author} {\bibfnamefont
  {S.}~\bibnamefont {Paschen}}, \bibinfo {author} {\bibfnamefont {M.~W.}\
  \bibnamefont {Haverkort}}, \bibinfo {author} {\bibfnamefont {L.~H.}\
  \bibnamefont {Tjeng}}, \ and\ \bibinfo {author} {\bibfnamefont
  {A.}~\bibnamefont {Severing}},\ }\href@noop {} {\  (\bibinfo {year}
  {2018}{\natexlab{b}})},\ \bibinfo {note} {{S}upplemental
  Material}\BibitemShut {NoStop}%
\bibitem [{\citenamefont {Willers}\ \emph {et~al.}(2012)\citenamefont
  {Willers}, \citenamefont {Strigari}, \citenamefont {Hiraoka}, \citenamefont
  {Cai}, \citenamefont {Haverkort}, \citenamefont {Tsuei}, \citenamefont
  {Liao}, \citenamefont {Seiro}, \citenamefont {Geibel}, \citenamefont
  {Steglich}, \citenamefont {Tjeng},\ and\ \citenamefont
  {Severing}}]{WillersPRL109}%
  \BibitemOpen
  \bibfield  {author} {\bibinfo {author} {\bibfnamefont {T.}~\bibnamefont
  {Willers}}, \bibinfo {author} {\bibfnamefont {F.}~\bibnamefont {Strigari}},
  \bibinfo {author} {\bibfnamefont {N.}~\bibnamefont {Hiraoka}}, \bibinfo
  {author} {\bibfnamefont {Y.~Q.}\ \bibnamefont {Cai}}, \bibinfo {author}
  {\bibfnamefont {M.~W.}\ \bibnamefont {Haverkort}}, \bibinfo {author}
  {\bibfnamefont {K.-D.}\ \bibnamefont {Tsuei}}, \bibinfo {author}
  {\bibfnamefont {Y.~F.}\ \bibnamefont {Liao}}, \bibinfo {author}
  {\bibfnamefont {S.}~\bibnamefont {Seiro}}, \bibinfo {author} {\bibfnamefont
  {C.}~\bibnamefont {Geibel}}, \bibinfo {author} {\bibfnamefont
  {F.}~\bibnamefont {Steglich}}, \bibinfo {author} {\bibfnamefont {L.~H.}\
  \bibnamefont {Tjeng}}, \ and\ \bibinfo {author} {\bibfnamefont
  {A.}~\bibnamefont {Severing}},\ }\href {\doibase
  10.1103/PhysRevLett.109.046401} {\bibfield  {journal} {\bibinfo  {journal}
  {Phys. Rev. Lett.}\ }\textbf {\bibinfo {volume} {109}},\ \bibinfo {pages}
  {046401} (\bibinfo {year} {2012})}\BibitemShut {NoStop}%
\bibitem [{\citenamefont {Prokofiev}\ and\ \citenamefont
  {Paschen}(2012)}]{Prokofiev_2012}%
  \BibitemOpen
  \bibfield  {author} {\bibinfo {author} {\bibfnamefont {A.}~\bibnamefont
  {Prokofiev}}\ and\ \bibinfo {author} {\bibfnamefont {S.}~\bibnamefont
  {Paschen}},\ }in\ \href@noop {} {\emph {\bibinfo {booktitle} {Modern Aspects
  of Bulk Crystal and Thin Film Preparation}}},\ Vol.~\bibinfo {volume} {11},\
  \bibinfo {editor} {edited by\ \bibinfo {editor} {\bibfnamefont
  {N.}~\bibnamefont {Kolesnikov}}\ and\ \bibinfo {editor} {\bibfnamefont
  {E.}~\bibnamefont {Borisenko}}}\ (\bibinfo  {publisher} {Intech Open Access
  Publisher},\ \bibinfo {year} {2012})\BibitemShut {NoStop}%
\bibitem [{\citenamefont {Wissgott}\ and\ \citenamefont
  {Held}(2016)}]{Wissgott2016}%
  \BibitemOpen
  \bibfield  {author} {\bibinfo {author} {\bibfnamefont {P.}~\bibnamefont
  {Wissgott}}\ and\ \bibinfo {author} {\bibfnamefont {K.}~\bibnamefont
  {Held}},\ }\href {\doibase 10.1140/epjb/e2015-60753-5} {\bibfield  {journal}
  {\bibinfo  {journal} {The European Physical Journal B}\ }\textbf {\bibinfo
  {volume} {89}},\ \bibinfo {pages} {5} (\bibinfo {year} {2016})}\BibitemShut
  {NoStop}%
\bibitem [{\citenamefont {Amorese}\ \emph {et~al.}(2014)\citenamefont
  {Amorese}, \citenamefont {Langini}, \citenamefont {Dellea}, \citenamefont
  {Kummer}, \citenamefont {Brookes}, \citenamefont {Braicovich},\ and\
  \citenamefont {Ghiringhelli}}]{Amorese_2018a}%
  \BibitemOpen
  \bibfield  {author} {\bibinfo {author} {\bibfnamefont {A.}~\bibnamefont
  {Amorese}}, \bibinfo {author} {\bibfnamefont {C.}~\bibnamefont {Langini}},
  \bibinfo {author} {\bibfnamefont {G.}~\bibnamefont {Dellea}}, \bibinfo
  {author} {\bibfnamefont {K.}~\bibnamefont {Kummer}}, \bibinfo {author}
  {\bibfnamefont {N.}~\bibnamefont {Brookes}}, \bibinfo {author} {\bibfnamefont
  {L.}~\bibnamefont {Braicovich}}, \ and\ \bibinfo {author} {\bibfnamefont
  {G.}~\bibnamefont {Ghiringhelli}},\ }\href@noop {} {} (\bibinfo {year}
  {2014}),\ \bibinfo {note} {submitted and arXiv:1410.1587}\BibitemShut
  {NoStop}%
\bibitem [{\citenamefont {Kummer}\ \emph {et~al.}(2017)\citenamefont {Kummer},
  \citenamefont {Tamborrino}, \citenamefont {Amorese}, \citenamefont {Minola},
  \citenamefont {Braicovich}, \citenamefont {Brookes},\ and\ \citenamefont
  {Ghiringhelli}}]{Kummer2017}%
  \BibitemOpen
  \bibfield  {author} {\bibinfo {author} {\bibfnamefont {K.}~\bibnamefont
  {Kummer}}, \bibinfo {author} {\bibfnamefont {A.}~\bibnamefont {Tamborrino}},
  \bibinfo {author} {\bibfnamefont {A.}~\bibnamefont {Amorese}}, \bibinfo
  {author} {\bibfnamefont {M.}~\bibnamefont {Minola}}, \bibinfo {author}
  {\bibfnamefont {L.}~\bibnamefont {Braicovich}}, \bibinfo {author}
  {\bibfnamefont {N.~B.}\ \bibnamefont {Brookes}}, \ and\ \bibinfo {author}
  {\bibfnamefont {G.}~\bibnamefont {Ghiringhelli}},\ }\href {\doibase
  10.1107/S1600577517000832} {\bibfield  {journal} {\bibinfo  {journal}
  {Journal of Synchrotron Radiation}\ }\textbf {\bibinfo {volume} {24}},\
  \bibinfo {pages} {531} (\bibinfo {year} {2017})}\BibitemShut {NoStop}%
\end{thebibliography}
\end{document}